\documentclass[showpacs,floatfix,twocolumn]{revtex4-1}
\usepackage{graphicx,psfrag,amsmath,amssymb,amsfonts,latexsym,color,epsf,dcolumn,graphpap}
\usepackage{subfig}
\usepackage{float}
\usepackage{lipsum}

\usepackage{tikz-cd} 
\usepackage{tikz}
\usetikzlibrary{shapes.geometric, arrows}

\definecolor{red}{rgb}{1,0,0}
\definecolor{blue}{rgb}{0,0,1}
\definecolor{skyblue}{rgb}{0,0,.5}
\definecolor{green}{rgb}{0,1,0}
\definecolor{orange}{cmyk}{0,.4,1,0}

\begin{document}
\title{Photon generation via dynamical Casimir effect in an optomechanical cavity as a closed quantum system}
\author{Nicol\'as F. Del Grosso, Fernando C.~Lombardo, Paula I.~Villar }
\affiliation{
Departamento de F\'\i sica {\it Juan Jos\'e
Giambiagi}, FCEyN UBA and IFIBA CONICET-UBA, Facultad de Ciencias Exactas y Naturales,
Ciudad Universitaria, Pabell\' on I, 1428 Buenos Aires, Argentina.}
\date{today}

\begin{abstract}
\noindent 
We present an analytical and numerical  analysis of the particle creation in an optomechanical cavity in parametric resonance. We treat both the electromagnetic field and the mirror as quantum degrees of freedom and study the dynamical evolution as a closed quantum system. We consider different initial states and investigate the spontaneous emission of photons from phonons in the mirror. We find that for initial phononic product states the evolution of the photon number can be described as a non-harmonic quantum oscillator, providing an useful tool so as to estimate the maximum and mean number of photons  produced for arbitrary high energies. The efficiency of this mechanism is further analyzed for a detuned cavity as well as the possibility of stimulating the photon production by adding some initial ones to the cavity. We also find relationships for the maximum and mean entanglement between the mirror and the wall in these states. Additionally we study coherent states for the motion of the mirror to connect this model with previous results from quantum field theory with a classical mirror. Finally we study thermal states of phonons in the wall and the equilibration process that leads to a stationary distribution.
\end{abstract}
\maketitle
\section{Introduction}\label{sec:intro}

One of the most striking features of quantum field theory (QFT) is that it predicts the production of particles from quantum vacuum. There are remarkable examples of the dynamical conversion of vacuum fluctuations into real particles: the Unruh radiation detected by a uniformly accelerating observer \cite{unruh}, the Hawking radiation originated from black holes \cite{Hawking1,Hawking2} and the Schwinger effect which produces pairs of electrons and positrons in the presence of a strong electromagnetic field (EM) \cite{Schwinger}. However, although there is very strong theoretical support for these effects, none has yet been observed experimentally.

A closely related phenomena is the dynamical Casimir effect (DCE) \cite{Moore, Lambrecht, Dodonov, Paraoanu, maianeto}, which consists in particle creation from time dependent external conditions. It has been first proposed in 1970 and its statement suggested that a Fabry-Perot cavity with one of its mirrors oscillating harmonically at twice the frequency of a mode field in the cavity, would lead to photon production from the vacuum \cite{Moore}. Later, it has been shown that a single mirror in free space subjected to non uniform acceleration would also produce photon radiation \cite{fulling}. However in all cases, the high accelerations required to produce photons were not attainable by physically moving massive mirrors. It has been a recent proposal,  the experimental suggestion that the DCE could be mimicked by tuning the boundary condition of the field. This idea lead  to the first experimental observation of the DCE \cite{Wilson}. In this experimental setup the cavity was replaced by a superconducting wave guide which ended with a superconducting quantum interference device (SQUID) and the boundary conditions were tuned by applying a time dependent magnetic flux through the SQUID. 
The observation of the DCE in superconducting circuits lacks, still, a fundamental part of the effect which is the conversion of mechanical energy into photons. Because of this the experimental realization has sometimes been called a simulation of the effect and a true observation in an optomechanical cavity is still awaiting. 

Optomechanical systems comprise an optical cavity formed by two mirrors one of which is free to vibrate. Practical optomechanical structures have been created in which the mirror can oscillate as fast as six billion times a second. However, this may not be quick enough; previous theoretical studies have shown that the mechanical oscillation frequency must be at least twice that one of the lowest energy cavity mode before DCE can be observed.  In a recent work \cite{prx}, authors treated both the cavity field and the moving mirror as quantum mechanical systems and noted the existence of vacuum Casimir-Rabi splittings for mirror frequencies lower than $\omega_{\rm c}$ . The analysis performed suggested that current optomechanical systems can be used to observe conversion of mechanical energy into light, which means that light emission from mechanical motion could be achieved in this kind of structures for lower frequencies. Moreover, recent developments in nano resonators technology together with higher Q cavities make the observation of the effect in optomechanical cavities accesible in the near future.
Another important feature suggested in \cite{prx}, has been the fact that the DCE could be analyzed in a more fundamental way with a time independent hamiltonian where an initial state with phonons in the wall would evolve to photons in the cavity. Further extensions of this model have been done in Refs. \cite{conversion, feasible}.

In this work, we follow the former idea and study the DCE as a closed system where both the EM field and the mirror are treated quantum mechanically. We investigate in detail the mechanism by which mechanical energy is converted into photons in parametric resonance ($\omega_{\rm m}=2\omega_{\rm c}$) in the weak coupling regime for different initial states.
\begin{figure*}
	\centering
	\begin{tikzpicture}
	\node[anchor=south west,inner sep=0] (image) at (0,0) {\includegraphics[width=0.55\linewidth]{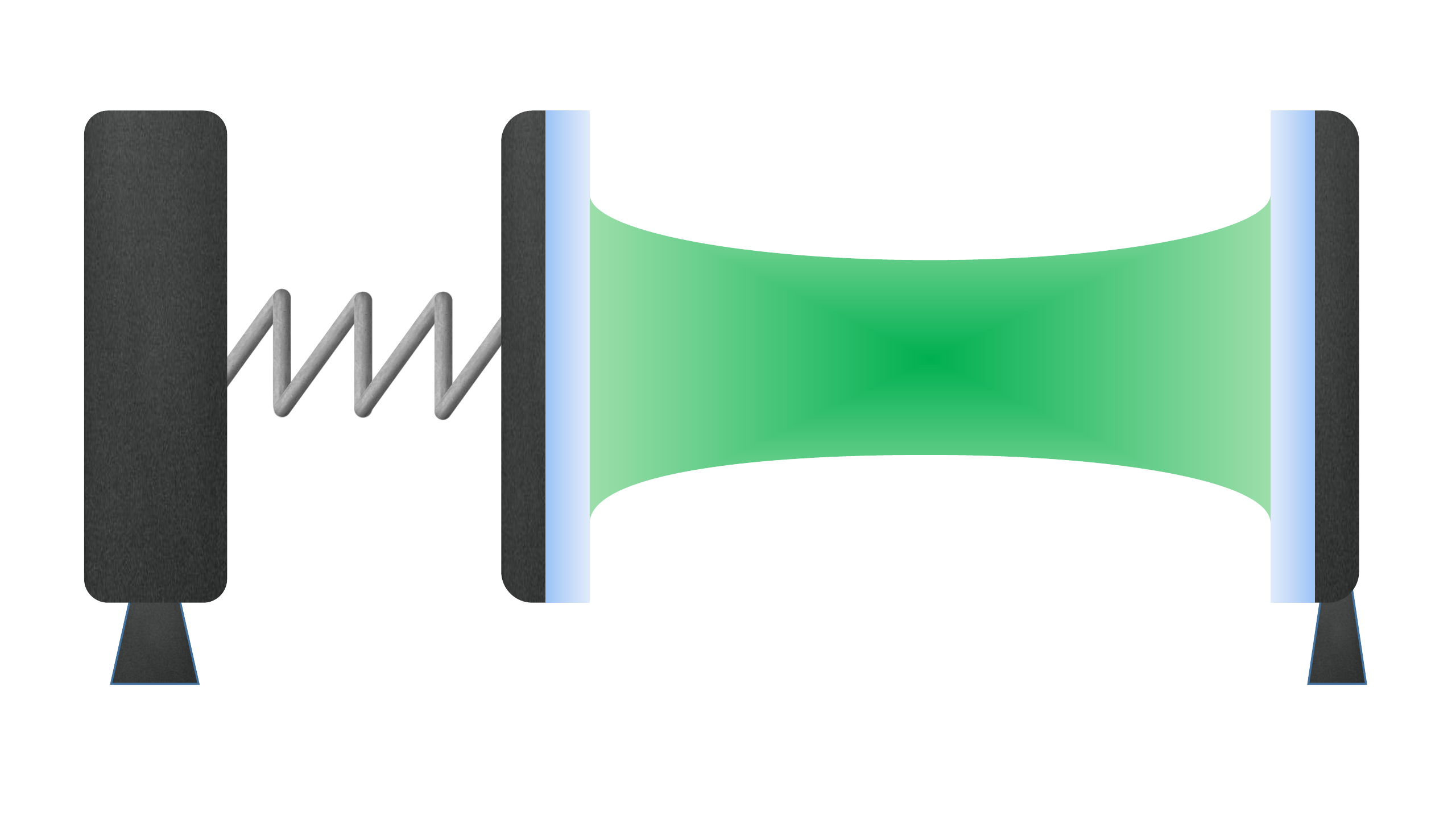}};
	\begin{scope}[x={(image.south east)},y={(image.north west)}]
	\node[] (wc) at (0.65,0.38){$\omega_{\rm c}$};
	\node[] (b) at (0.65,0.75){$\hat{a}$};
	\node[] (wm) at (0.25,0.43){$\omega_{\rm m}$};
	\node[] (a) at (0.25,0.7){$\hat{b}$};
	\draw[line width=0.3mm,<->] (0.33,0.2) -- (0.43,0.2);
	\node[] (x) at (0.38,0.1){$x(t)$};
	\end{scope}
	\end{tikzpicture}
	\caption{\label{fig:sistema} We present the model studied: a cavity with a movable end mirror (and the other end fixed). The moving mirror is subjected to a harmonic potential analogous to a spring with frequency $\omega_{\rm m}$  attached to a fixed wall. The oscillations of the spring are quantized with bosonic operator $\hat{b}$ corresponding to phonons. Inside the cavity, light is produced as a result of the dynamical Casimir interaction in one electromagnetic mode with frequency $\omega_{\rm c}$ and bosonic operator $\hat{a}$.}
\end{figure*}
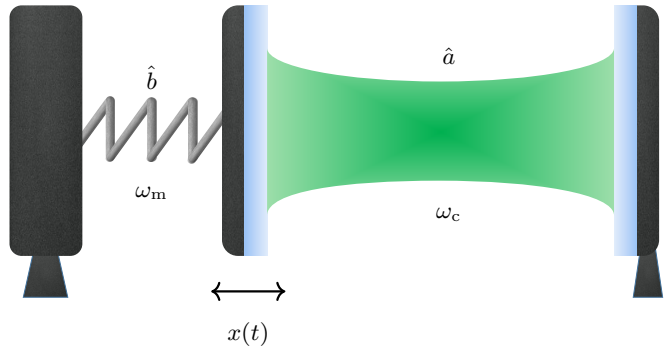
The paper is organized as follows.
In Section II we describe the model for an optomechanical cavity where both degrees of freedom, the field and mirror motion, are described quantum mechanically and commentate on possible refinements. In Sec. III we numerically and analytically study  the time evolution of product states between the mirror and the field. We focus on the efficiency of photon production and the development of entanglement for growing energy states. We also study how this efficiency for different situations:  an initial state with some photons already in the cavity and and off-parametric resonance. In Sec. IV, we perform a numerical analysis in order to make a connection between the system treated a closed quantum system and the quantum field theory model with a semiclassical wall. Thus, we consider an initial state with a coherent motion of the mirror.  Sec. V is dedicated to the numerical study of the equilibration process resulting from an initially hot mirror and a vacuum cavity. Finally, in Sec. VI, we summarize our results and present the conclusions of our work.


\section{Model}
Herein, we shall describe the optomechanical system in which we study the DCE. 
We shall consider a massless scalar field $\phi(x,t)$ inside a cavity $[0, x(t)]$ with a mobile wall that obeys the wave equation
\begin{equation}\label{eqh}
\frac{\partial^2 \phi(x,t)}{\partial t^2}=\frac{\partial^2 \phi(x,t)}{\partial x^2}
\end{equation}
and satisfies Dirichlet boundary conditions
\begin{equation}\label{eqh}
\phi(0,t)=\phi(x(t),t)=0.
\end{equation}

By subjecting the mobile wall to a harmonic potencial and letting it interact with the field through radiation pressure, we can obtain a hamiltonian description of the system \cite{law}. After keeping only one field mode and applying a canonical quantization, the field will be described by photons with bosonic operator $\hat{a}$ and the wall oscillations by phonons with bosonic operator $\hat{b}$. Mathematically, this results in the following hamiltonian
\begin{equation}\label{eqh}
H=H_0+V_{\rm om}+V_{\rm DCE},
\end{equation}
where 
\begin{equation}\label{eqh}
H_0=\hbar\omega_{\rm c}\hat{N_{ a}}+\hbar\omega_{\rm m}\hat{N_{ b}}
\end{equation}
is the free hamiltonian, composed by the photon and phonon number operators $\hat{N_{\rm a}}=\hat{a}^{\dagger}\hat{a}$ and $\hat{N_{\rm b}}=\hat{b}^{\dagger}\hat{b}$ respectively, while
\begin{equation}\label{eqh}
V_{\rm om}=g\hbar\ \hat{N_{ a}}(\hat{b}+\hat{b}^{\dagger})
\end{equation}
is the optomechanical interaction between the mirror and field with coupling strength $g$, and
\begin{equation}\label{eqh}
V_{\rm DCE}=\frac{g\hbar}{2}(\hat{a}^{2}+\hat{a}^{\dagger2})(\hat{b}+\hat{b}^{\dagger})
\end{equation}
is the dynamical Casimir effect interaction. 

The free hamiltonian $H_0$ has eigenstates given by the product basis 
\begin{equation}
	|n,k\rangle=\hat{a}^{\dagger n}\hat{b}^{\dagger k}|0,0\rangle
\end{equation}
formed by $n$ photons and $k$ phonons.

We can easily see that the hamiltonian $H$ captures some of the key features expected from the DCE. Namely that, since it does not commute with the photon number operator $\hat{N}_{a}$, it is possible to start with the cavity in the vacuum state and produce photons from phonons in the wall. Another expected feature is the production of light in photon pairs which corresponds to the term $\hat{a}^{\dagger2}\hat{b}$ that converts a phonon into a photon pair. In fact, if we decompose our Hilbert space as 
\begin{equation}
	\mathcal{H}=\mathcal{E}\oplus \mathcal{O},
\end{equation}
where $\mathcal{E}$ is the subspace of states with an even number of photons and $\mathcal{O}$ the one with an odd number of photons; we can see that the hamiltonian leaves invariant the subspaces: $H\mathcal{E}\subseteq\mathcal{E}$ and  $H\mathcal{O}\subseteq\mathcal{O}$. This means that given $|e\rangle \in \mathcal{E}$ $\rightarrow$ $|e(t)\rangle=e^{-iHt/\hbar}|e\rangle \in \mathcal{E}$ and thus, given an initial state in one of these subspaces, the time evolved state will remain in that same subspace.

This model with only 1 photon mode is a good approximation as long as the other modes are not significantly excited by the phonons. Previous results \cite{crocce, squid1,squid2, 2walls,pre} show that this happens when the frequencies of the modes are not equally spaced which happens for a massive field in 1 dimension or a massless field in 2 or more dimensions.
In addition, we must say that we are considering only 1 polarization state of the electromagnetic field  in this model. Considering both of them only duplicates the problem, with the hamiltonian being given by
\begin{align}\label{eqh2}
H&=\hbar\omega_{\rm c}\hat{N}_{a,x}+\hbar\omega_{\rm m}\hat{N}_{b}+g\hbar\ N_{a,x}(\hat{b}+\hat{b}^{\dagger}) \nonumber\\
&+\frac{g\hbar}{2}(\hat{a}_{\rm x}^{2}+\hat{a}_{\rm x}^{\dagger2})(\hat{b}+\hat{b}^{\dagger})+\hbar\omega_{\rm c}\hat{N}_{a,y} \nonumber\\
&+g\hbar\ \hat{N}_{a, y}(\hat{b}+\hat{b}^{\dagger})+\frac{g\hbar}{2}(\hat{a}_{\rm y}^{2}+\hat{a}_{\rm y}^{\dagger2})(\hat{b}+\hat{b}^{\dagger}).
\end{align}
However, the $x$ and $y$ polarization states, $\hat{a}_{\rm x}^{\dagger n}|0\rangle$ and $\hat{a}_{\rm y}^{\dagger n}|0\rangle$, are not eigenstates of the true electromagnetic hamiltonian since they do not commute with the helicity operator, that is they do not have a well defined spin. The eigenstates of the EM hamiltonian are given by $\hat{a}_{\uparrow}^{\dagger n}|0\rangle$ and $\hat{a}_{\downarrow}^{\dagger n}|0\rangle$ with $\hat{a}_{\uparrow}=\hat{a}_{\rm x}+i\hat{a}_{\rm y}$ and $\hat{a}_{\downarrow}=\hat{a}_{\rm x}-i\hat{a}_{\rm y}$. Writing the former hamiltonian with these operators shields
\begin{align}
H&=\hbar\omega_{\rm c}(\hat{N}_{a,\uparrow}+\hat{N}_{a,\downarrow})+\hbar\omega_{\rm m}\hat{N}_{b} \\
&+g\hbar\ (\hat{N}_{a,\uparrow}+\hat{N}_{a,\downarrow})(\hat{b}+\hat{b}^{\dagger})+\frac{g\hbar}{2}(\hat{a}_{\uparrow}\hat{a}_{\downarrow}+\hat{a}_{\uparrow}^{\dagger}\hat{a}_{\downarrow}^{\dagger})(\hat{b}+\hat{b}^{\dagger}). \nonumber
\end{align}
We can see that the only difference with our model is that the pairs of photons would be produced in an entangled EPR state $(|\uparrow\downarrow\rangle+|\downarrow\uparrow\rangle)/\sqrt{2}$. A final comment about the model is that it would be possible to add a coherent mechanical drive of the mirror by adding to the hamiltonian the time dependent term as
\begin{align}
V_{\text{d}}=f(t)(\hat{b}+\hat{b}^\dagger),
\end{align}
with $f(t)$ proportional to the force applied to the mirror. However, in this work we shall focus on the interconversion of phonons to photons through the DCE interaction as a closed quantum system.

\begin{figure*}
	\subfloat[\label{sfig:prod_vm}]{%
		\includegraphics[width=.45\linewidth]{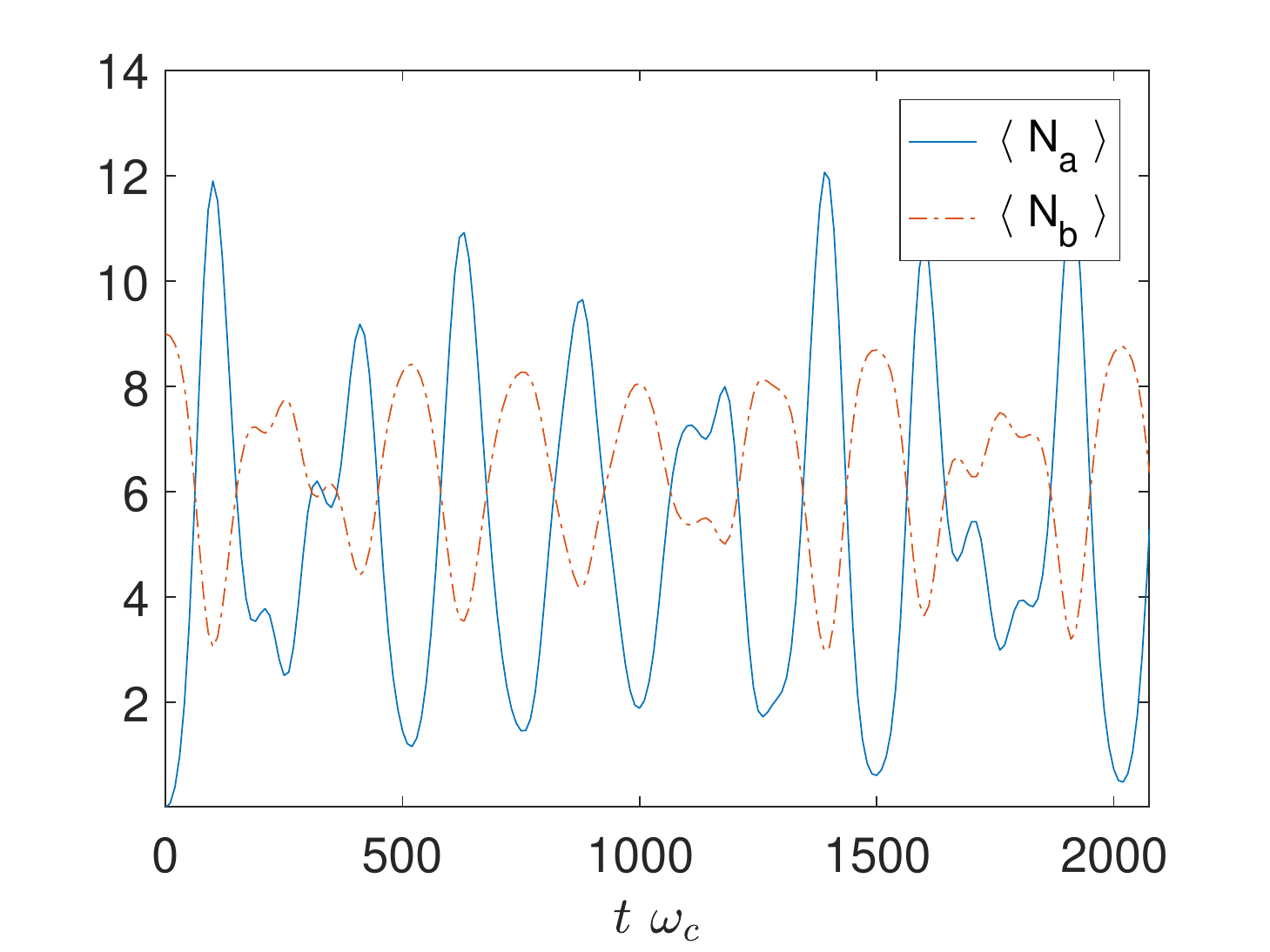}%
	}\hfill
	\subfloat[\label{sfig:prod_eta}]{%
		\includegraphics[width=.45\linewidth]{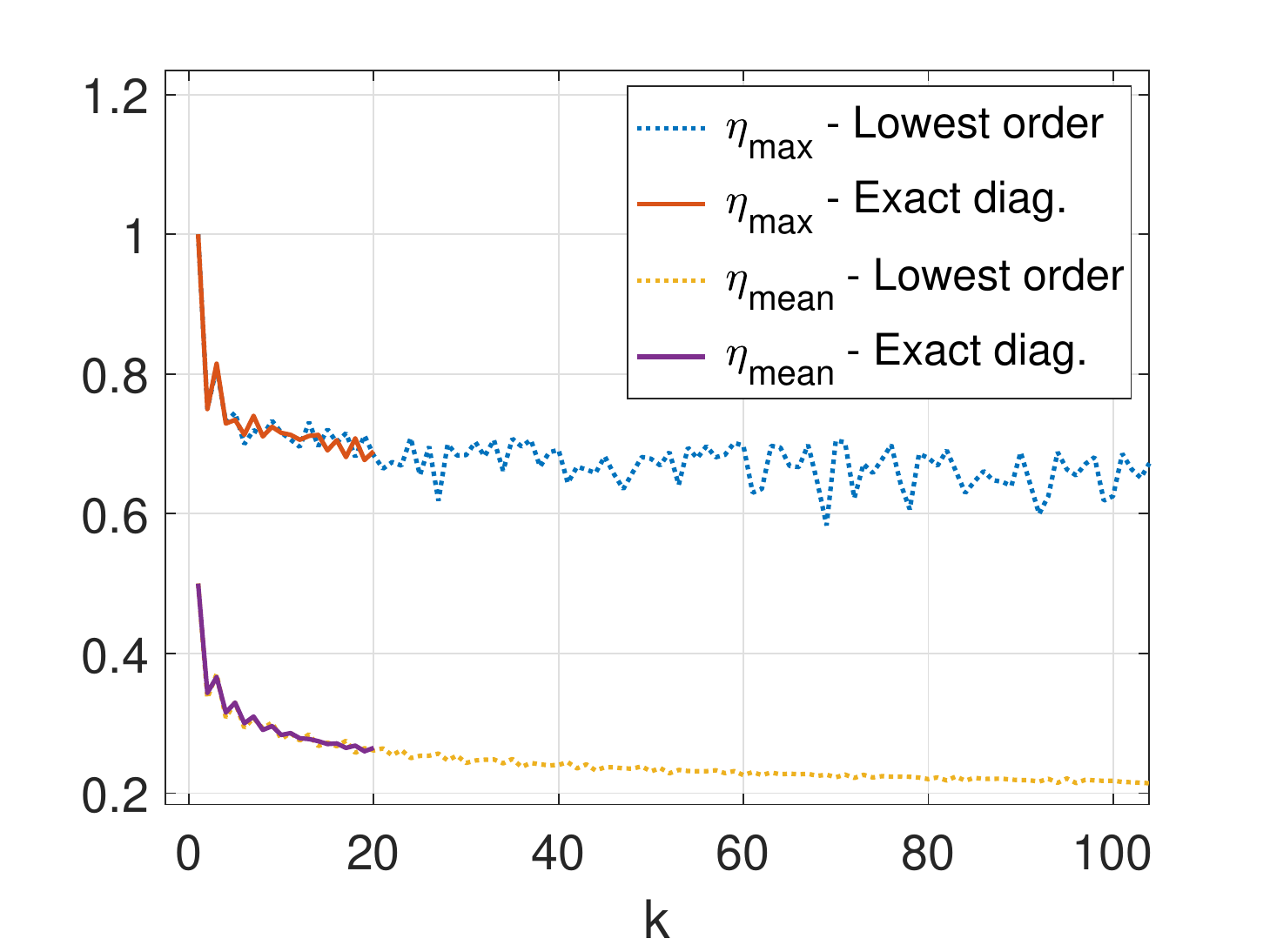}%
	}\hfill
	\subfloat[\label{sfig:prod_s_t}]{%
		\includegraphics[width=.45\linewidth]{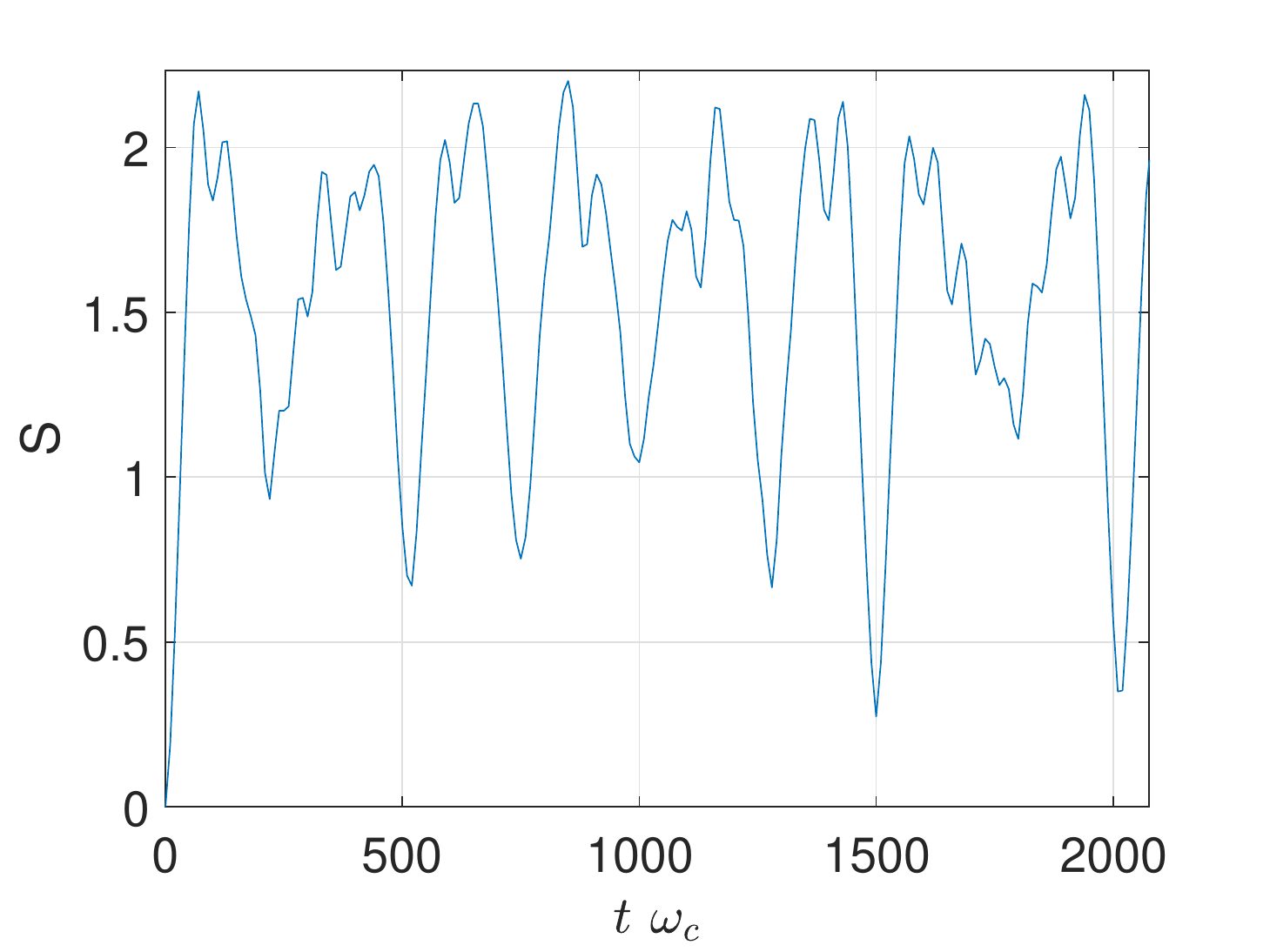}%
	}\hfill
	\subfloat[\label{sfig:prod_s_k}]{%
		\includegraphics[width=.45\linewidth]{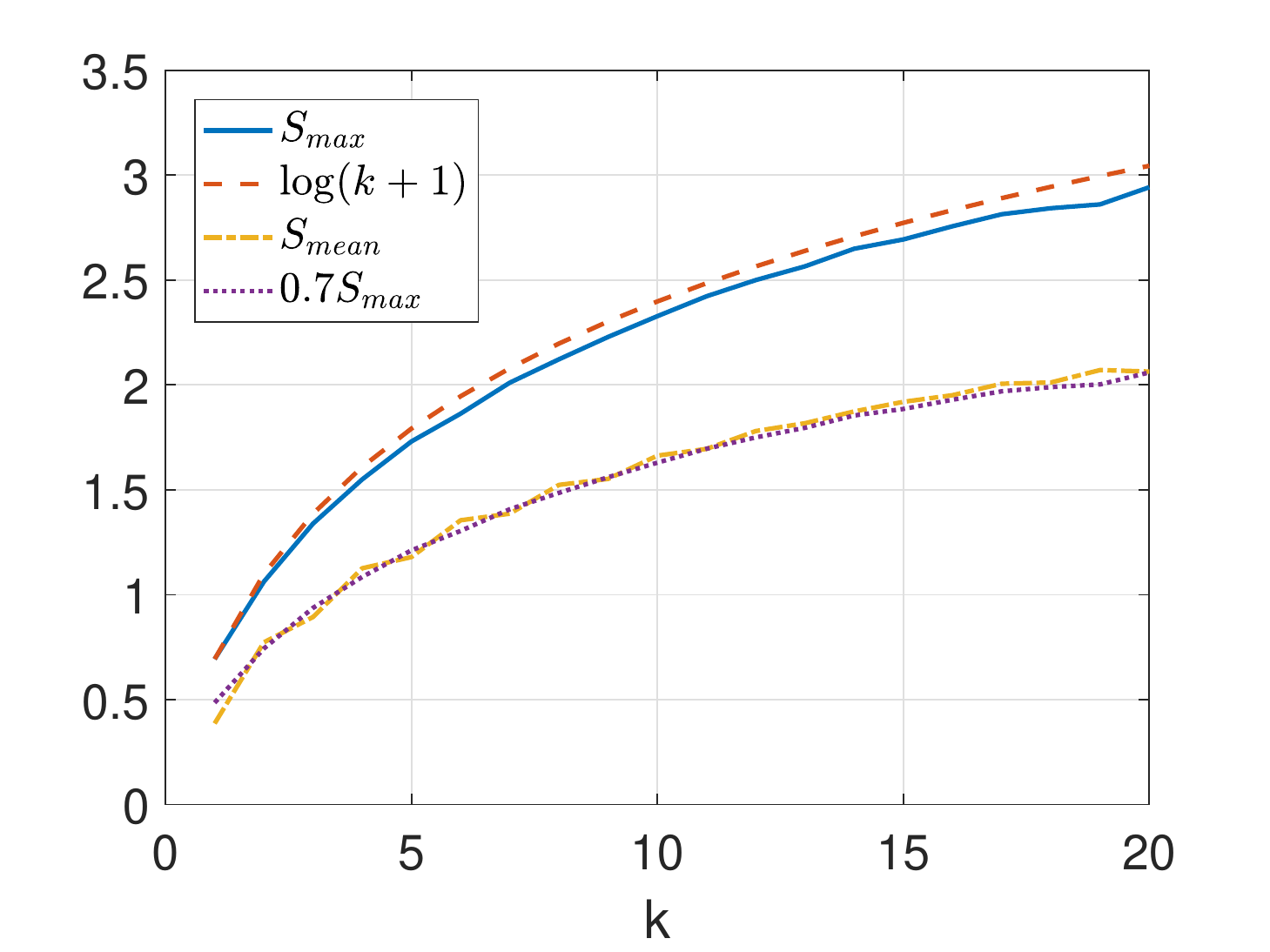}%
	}
	\caption{a) Time evolution for the mean photon and phonon numbers. They oscillate erratically in time for the generic state $|0,9\rangle$. b) Both the maximum and mean efficiency of mechanical to photon energy conversion can be seen to converge for a large number of initial phonons. We show both magnitudes as a function of the initial phonon number obtained by exact diagonalization of the hamiltonian and by using lowest order perturbation theory. c) The entanglement entropy as a function of time for the generic state $|0,9\rangle$. d) Both the maximum and mean entanglement entropy show a logarithmic dependence with the initial number of phonons. In all cases, results were obtained for $g=0.01$ and parametric resonance.}
	\label{fig:tensor_vm}
\end{figure*}

\section{Product states}

The DCE is expected to produce an exponential increase of energy for a coherent external driving of the wall and so in this section we will study the dynamics of the system for initial states with growing energy. Precisely, we will consider initial states of the form 
\begin{equation}
	|\psi_0\rangle=|0,k\rangle
\end{equation}
with the system in parametric resonance, that is $\omega_{\rm m}=2\omega_{\rm c}$, and weakly coupled ($g/\omega_c\ll1$). These are eigenstates of the free hamiltonian and belong to the degenerate subspace $D_{\rm k}$ generated by the basis $\{ |2n,k-n\rangle \}_{0\leq n \leq k}$. We can solve the dynamics in the weak coupling regime by using perturbation theory to lowest order, which corresponds to diagonalizing the restriction of $V_{\text{DCE}}$ to $D_{\rm k}$ given by
\begin{equation}
	V_{\text{DCE}}\vert_{D_{\rm k}}=\frac{g\hbar}{2}\left(\begin{array}{ccccc}
	0 & v_{1} & 0 & 0 & 0\\
	v_{1} & 0 & v_{2} & 0 & 0\\
	0 & v_{2} & 0 & ... & 0\\
	0 & 0 & ... & 0 & v_{\rm k}\\
	0 & 0 & 0 & v_{k} & 0
	\end{array}\right),
\end{equation}
with  $v_{\rm j}=\sqrt{2j(2j-1)(k+1-j)}$.

For the case of $k=1$, it has been done in \cite{prx}, where they showed that the state is given by 
\begin{equation}
	|\psi(t)\rangle=\cos(\Omega t)|0,1\rangle+i\sin(\Omega t)|2,0\rangle. 
\end{equation}
Thus, it is possible to convert all the mechanical energy of the phonon mode into electromagnetic energy for $t=\pi/(2\Omega)$. Herein, we shall study this interconversion of energy in more detail for a variety of different initial states.

In Fig. \ref{sfig:prod_vm} we present the time evolution of the mean number of photons $\langle \hat{N}_{a}\rangle$ and phonons $\langle \hat{N}_{b}\rangle$ for the representative initial state $|0,9\rangle$. Both of these magnitudes oscillate in time between a maximum value and 0, being  correlated via conservation of energy $\langle \hat{N}_{a}\rangle(t)+2\langle \hat{N}_{b}\rangle(t)=2k$,  in the weak coupling regime. An interesting feature occurs: even though the maximum value of $\langle \hat{N}_{a}\rangle(t)$ grows lineally with the initial number of phonons, it is always less than the $2k$ (allowed by the conservation of energy), except for $k=1$. In fact, if we define the maximum efficiency of converting mechanical energy into electromagnetic energy as 
\begin{equation}
\eta_{\text{max}}:=\max_{t}\frac{E_{a}(t)-E_{a}(0)}{E_{b}(0)},
\end{equation}
where $E_{a}(t)=\hbar\omega_{\rm c} \langle \hat{N}_{a}\rangle(t)$ and $E_{b}(t)=\hbar\omega_{\rm m} \langle \hat{N}_{b}\rangle(t)$ are the photon and phonon energies respectively and analogously $\eta_{\text{mean}}$; then, $\eta_{\text{max}}$ starts in $100\%$ for only $1$ initial phonon and has an asymptotic value as $k\to \infty$ of around $70\%$ (as we can see in Fig. \ref{sfig:prod_eta}). This asymmetry derives from the $V_{\text{DCE}}$ interaction which distinguishes between photons and phonons.

It is possible to study the entanglement between the EM field inside the cavity and the wall.To this end, we use the entanglement entropy $S=-tr(\hat{\rho}_{a}\log(\hat{\rho}_{a}))$, with $\hat{\rho}_{a}$ the reduced density matrix of photons. In Fig. \ref{sfig:prod_s_t} we can note that this magnitude oscillates in time between 0 and $\log(k+1)$,  which means that we can approximately recover the initial state in a short timescale. On the other hand, the behavior of  $S_{\rm max}\approx\log(k+1)$ can be well understood by using perturbation theory. As the initial state $|0,k\rangle$ belongs to $D_{\rm k}$, the time evolved state $|\psi\rangle(t)$ will also be in $D_{\rm k}$, which means its entanglement entropy is bounded by the maximally entangled stated
\begin{equation}
	|\psi\rangle=\sum_{n=0}^{k} \frac{1}{\sqrt{k+1}} |2n,k-n\rangle 
\end{equation}
whose entropy is  $S_{\rm max}=\log(k+1)$. Therefore, as time passes, an initially pure state evolves into a maximally entangled state (in the allowed subspace) and approximately returns to itself.
This large amount of entanglement is not an exceptional value along the time evolution but actually it is the rule. In fact, the mean value of the entanglement entropy coincides to a high degree of accuracy with $70\%$ of the maximum. We might speculate that by externally driving the system, we will increase the energy of phonons exponentially. This, if  combined with the logarithmic behavior found for the entanglement, would shield a linear increase of the entanglement entropy with time (Fig. \ref{sfig:prod_vm}).

\subsection{Stimulated and Inhibited photon creation}

\begin{figure}
 \centering
		\includegraphics[scale=0.6]{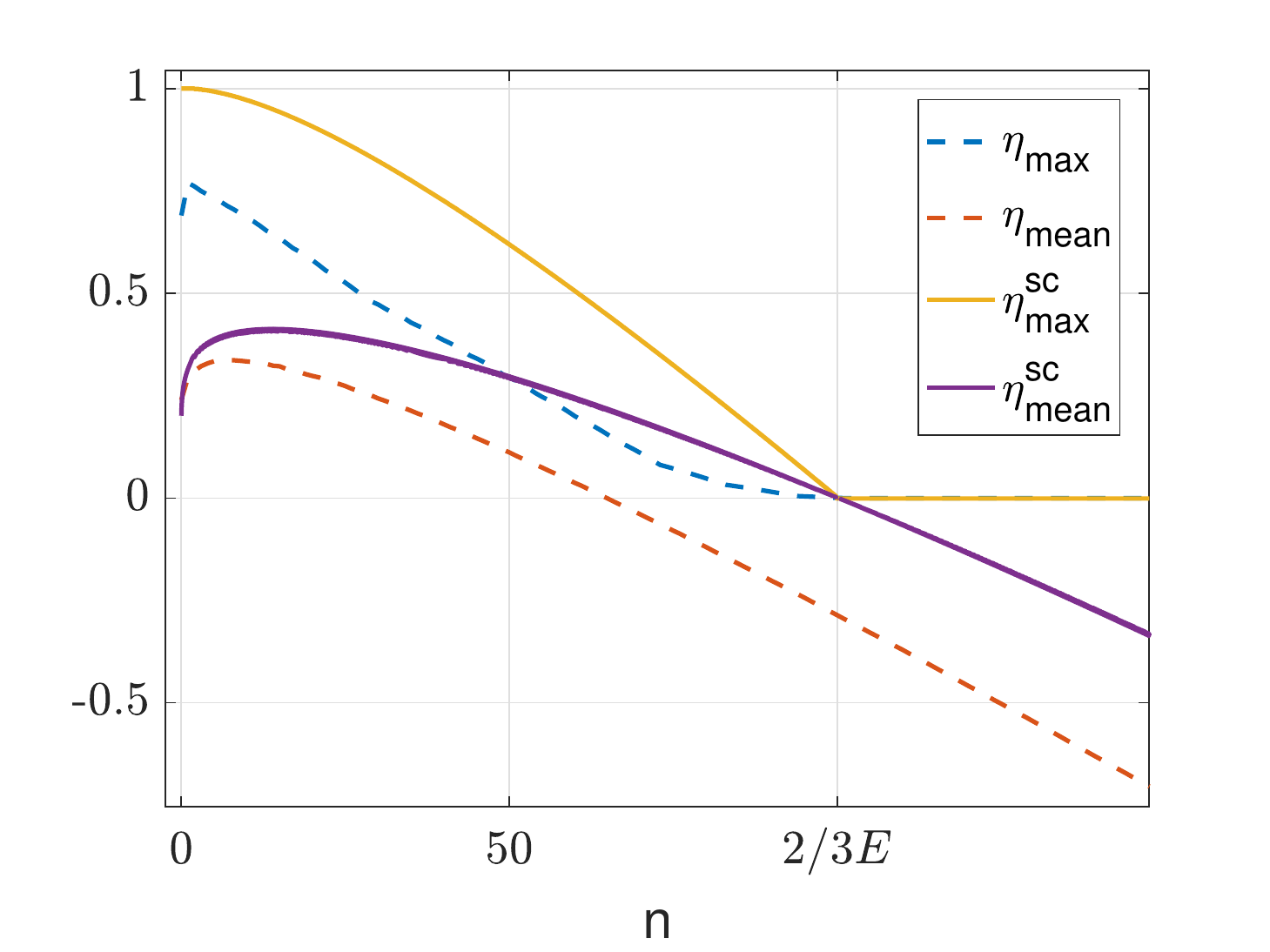}
		\caption{ Efficiency of energy conversion as a function of the pairs of photons for an initial state of the form $|2n,50\rangle$. The exact magnitudes obtained by numerical diagonalization $\eta_{\text{max}}$ and $\eta_{\text{mean}}$ are compared with the semiclassical approximations $\eta_{\text{max}}^{\rm sc}$ and $\eta_{\text{max}}^{\rm sc}$.}
		\label{fig:tensor_eta_n}
\end{figure}

In the previous section we have analyzed the dynamics and efficiency of photon production for states with an initially empty cavity. The natural question that might consequently arise is about what happens if we start with already some photons  in the cavity and phonons in the wall. Would this stimulate the production of even more photons increasing the efficiency or do they instead inhibit the conversion?
To answer the question,  we consider an initial state of the form $|2n,k_0\rangle$ and study the maximum and mean efficiency $\eta_{\rm mean}$ of photon creation (defined analogously to $\eta_{\rm max}$) for a fixed number of initial phonons $k_0$ but varying the number of photons $n$. We can say that we take the initial photons as catalyst and analyze how many new ones we can produce from the initial phonons.
In Fig. \ref{fig:tensor_eta_n}, we show the efficiency of conversion as a function of the pairs of photons for an initial state. It can be seen that it is actually better to have a few photons already in the initial state in order to increase the efficiency. However it must be noted  that adding too many can inhibit the production of more photons. Both measures of efficiency agree on this fact, even though  $\eta_{\rm max}$ peaks for just 1 pair of photons, while $\eta_{\rm mean}$ does it for around $10\%$ of the initial phonons. After reaching these peaks, both efficiencies decrease almost linearly with growing number of initial photons. The maximum efficiency becomes 0 when the photons exceed $2/3$ of the total energy, while the mean efficiency becomes negative when the photons are larger than $56\%$ of the energy and then continues to decrease. The negative values of the mean efficiency reflect the fact that the initial photons stop acting as a catalyst and, instead, start to be consumed to produce phonons; while the maximum efficiency (being non negative by definition) falls to cero since the maximum number of photons is actually the initial one. This occurs because if the initial state has $0.56k<n<0.66k$,  it subsequently reaches an equilibrium between the number of photons and phonons where neither is created nor destroyed. 
We can understand this fact through a first approach, by rewriting the hamiltonian in position space
\begin{align}
H_{\rm ps}   &=\hbar\omega_c(X^2+P_{\rm X}^2)+\hbar\omega_m(Y^2+P_{\rm Y}^2)-2g\hbar X^2Y
\end{align}
with
\begin{align}
	X   &=\hat{a}+i\hat{a}^\dagger  \\
	P_{\rm X} &=\hat{a}-i\hat{a}^\dagger  \\
	Y   &=-\hat{b}-i\hat{b}^\dagger  \\
	P_{\rm Y} &=-\hat{b}+i\hat{b}^\dagger.  
\end{align}
We note that the classical system associated to this hamiltonian has a non trivial equilibrium position at $(X_{\rm EQ}, Y_{\rm EQ})=(\sqrt{\omega_{\rm m} \omega_{\rm c}}/(g\sqrt{2}), \omega_{\rm c}/(2g))$. In parametric resonance, this equilibrium corresponds to $2/3$ of the total energy being in the photon mode  which explains why $\eta_{\rm max}\simeq0$ around that point. Likewise, we would like to explain how this fact arises quantum mechanically. Hence, we can perform a RWA approximation of the original hamiltonian and discard the fast oscillating terms to get a  simpler one $H_{\rm RWA}=H_0+V_{\rm RWA}$ with
\begin{align}
V_{\rm RWA}   &=\frac{g\hbar}{2}(\hat{a}^{2}\hat{b}^{\dagger}+\hat{a}^{\dagger2}\hat{b})
\end{align}
and $H_0=\hbar \omega_{\rm c} \hat{N}_{a}+2\hbar \omega_{\rm c} \hat{N}_{b}$ the free hamiltonian.
It is possible to see that, for parametric resonance, $[H_{\rm RWA} , H_0]=0 $ and as this hamiltonian commutes with the free one, it conserves the sum of the modes energies. Then, using the Heisenberg equation for the slowly changing variables $\bar{\hat{b}}=e^{-i\omega_{m}t}\hat{b}$ and $\bar{\hat{a}}^2=e^{-i\omega_{m}t}\hat{a}^2$, we have
\begin{align}
\frac{d\hat{N}_{a}}{dt}&=\frac{1}{i\hbar}[\hat{N}_{a},H_{\rm RWA}]=\frac{g}{2}\frac{2}{i}\left(\bar{\hat{a}}^{\dagger2}\hbar-\bar{\hat{a}}^{2}\bar{\hat{b}}^{\dagger}\right)\\
\frac{d\bar{\hat{b}}}{dt} &=t \frac{1}{i\hbar}[\bar{\hat{b}},V_{\rm RWA}]=\frac{g}{2}\frac{1}{i}\bar{\hat{a}}^2\\
\frac{d\bar{\hat{a}}^2}{dt} &= \frac{1}{i\hbar}[\bar{\hat{a}}^2,V_{\rm RWA}]=\frac{g}{2}\left(\frac{2}{i}\bar{\hat{b}}+\frac{4}{i}\hat{N}_{\rm a}\bar{\hat{b}}\right).
\end{align}
Once more, we can derive the equation for $\hat{N}_{\rm a}$ with respect to time. By combining it with the other two  equations, we get
\begin{align}
\frac{d^2\hat{N}_{a}}{dt^2}&=\left(\frac{g}{2}\right)^24(2\hat{N}_{b}+\hat{N}_{a})+\left(\frac{g}{2}\right)^2 4\hat{N}_{a}(4\hat{N}_{b}-\hat{N}_{a}) \nonumber\\
&=g^2\frac{H_0}{\hbar \omega_{\rm c}}+g^2 \hat{N}_{a}\left(2\frac{H_0}{\hbar \omega_{\rm c}}-3 \hat{N}_{a}\right),
\end{align}
where we have used that $\hat{N}_{b}=(H_0-\hat{N}_{a})/2$. Since we are interested in the evolution of a product state $|\psi_0 \rangle$ in subspace $D_{\rm k}$, 
 we can just call $\langle H_0\rangle=E$ and write
\begin{align}\label{eq:semiclas}
\frac{d^2\hat{N}_{a}}{dt^2}&=g^2\frac{E}{\hbar \omega_{\rm c}}+g^2 \hat{N}_{a}\left(2\frac{E}{\hbar \omega_{\rm c}}-3 \hat{N}_{a}\right),
\end{align}
using that $H_0\vert_{D_{\rm k}}=E\ Id$. Finally, noting that this equation can be written as $\frac{d^2\hat{N}_{a}}{dt^2}=-V'(\hat{N}_{a})$, with 
\begin{equation}
	V( \hat{N}_{a})=g^2\left(\hat{N}_{a}^3-\frac{E}{\hbar \omega_{\rm c}} \hat{N}_{a}^2\right)-g^2\frac{E}{\hbar \omega_{\rm c}} \hat{N}_{a};
\end{equation}
we can see that the evolution of the number of photons corresponds to a non-harmonic quantum oscillator.
Hence, we can take expectation values and, making a semiclassical approximation $\langle \hat{N}_{a}\hat{N}_{ b}\rangle \simeq \langle \hat{N}_{a}\rangle\langle \hat{N}_{b}\rangle$, $\langle \hat{N}_{a}^{2}\rangle \simeq \langle \hat{N}_{a}\rangle^{2}$, we obtain the following estimation
\begin{figure}
	\centering
	\includegraphics[scale=0.6]{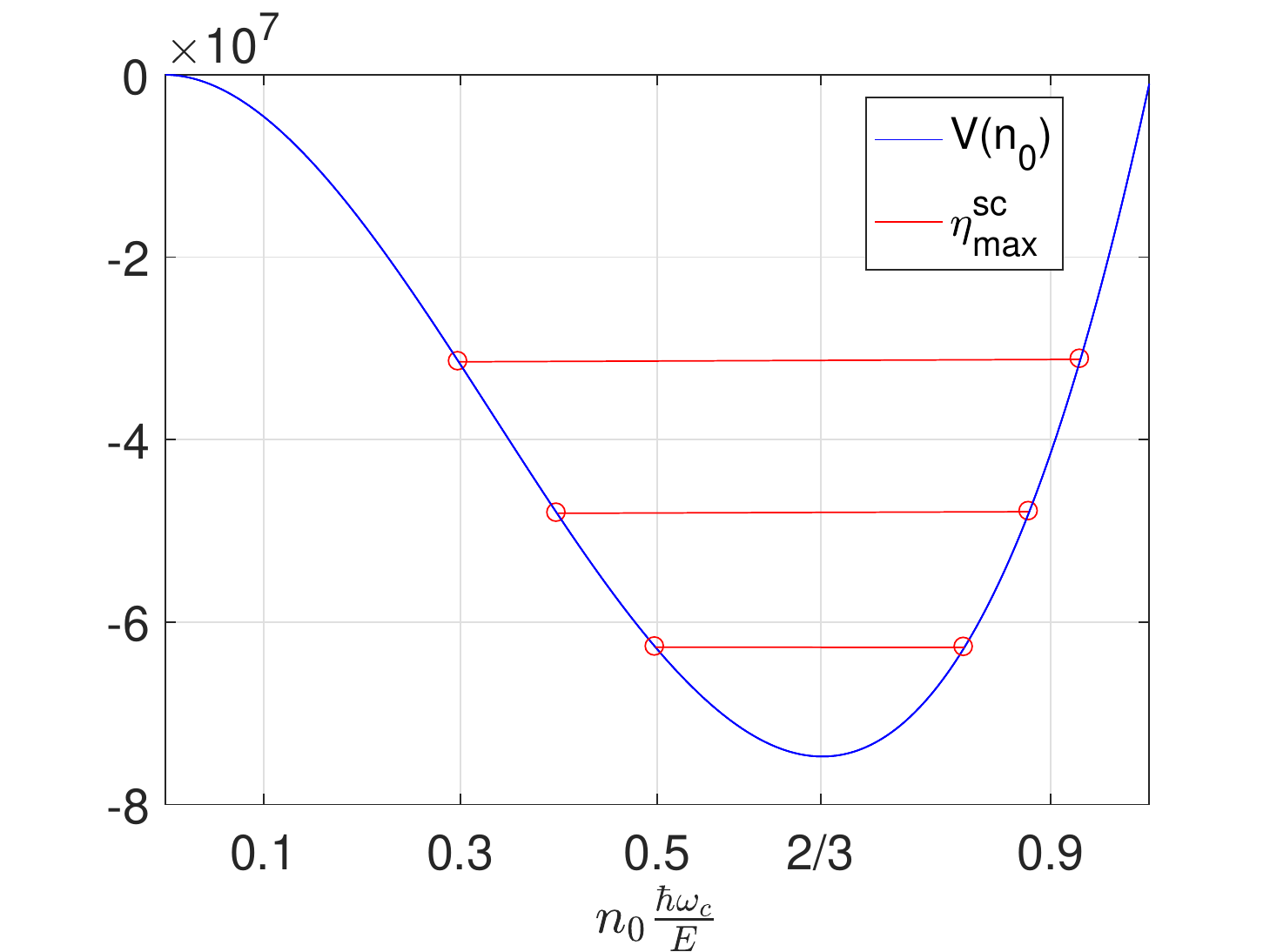}
	\caption{ Effective potential for $E/(\hbar\omega_{\rm c})=500$ and $g=1$. A product state with a mean value of $n_0$ initial photons evolves to another with a maximum of $n_1$ photons, where $n_1$ is given by $V(n_0)=V(n_1)$. The difference between $n_0$ and $n_1$ is proportional to $\eta_{\text{max}}$.}
	\label{fig:pozo}
\end{figure}
\begin{align}\label{eq:semiclas2}
\frac{d^{2}\langle \hat{N}_{a}\rangle}{dt^{2}} &=g^2\frac{E}{\hbar \omega_{\rm c}}+g^2\langle \hat{N}_{a}\rangle\left(2\frac{E}{\hbar \omega_{\rm c}}-3\langle \hat{N}_{a}\rangle\right).
\end{align}
This last differential equation captures the dynamics that converts phonons to photons in a simply approximate way. We can think of $\langle \hat{N}_{a}\rangle$ as the position of a particle moving in the potential $V(\langle \hat{N}_{a}\rangle)$ which, for $E\gg1$, again has a minimum in $\hbar\omega_{\rm c}\langle \hat{N}_{a}\rangle=2/3 E$ and satisfies that $V(0)=V(E/(\hbar \omega_{c}))=0$. This tells us that the solutions to this equation with initial conditions $(\langle \hat{N}_{a}\rangle(0), d\langle \hat{N}_{a}\rangle/dt(0))=(n_0,0)$ are oscillations $\langle \hat{N}_{a}\rangle(t)$ between $n_0$ and $n_1$ such that $V(n_1)=V(n_0)$, as illustrated in Fig. \ref{fig:pozo}. Using this we can calculate the semiclassical efficiency $\eta_{\text{max}}^{\rm sc}$, which reproduces quite well the main features of the exact solution $\eta_{\text{max}}$ shown in Fig. \ref{fig:tensor_eta_n}. The difference among the curves  originate in quantum correlations that produce some sort of dissipation. The semiclassical result sets then a useful bound on the efficiency of photon production.

Finally, we can mention that Eq. (\ref{eq:semiclas}) can also be used to understand the convergence  of the efficiency for a large number of phonons (Fig. \ref{sfig:prod_eta}). 
Defining $\hat{\eta}:=\hbar \omega_{\rm c}\hat{N}_{\rm a}/E$, we can use Eq. (\ref{eq:semiclas}) to get a differential equation for $\hat{\eta}$ which, by assuming $E\gg1$
\begin{equation}
	\frac{d^2\hat{\eta}}{dt^2}=\frac{g^2E}{\hbar \omega_{\rm c}} \hat{\eta}\left(2-3 \hat{\eta} \right).
\end{equation}
the result  is a differential equation for a non-harmonic quantum oscillator moving in the potential $V_{\text{eff}}(\hat{\eta})=g^2E/(\hbar \omega_{\rm c})(\hat{\eta}^3-\hat{\eta}^2)$. Therefore the solution does not depend on the energy. This is a very useful result since it allows us to quickly predict the maximum number of photons that will be generated for a given state. Previously, given the state $|248, 251\rangle$, we were forced to diagonalize a $375\times375$ matrix so as to perform perturbation theory and thus obtain the number of
photons generated. However, by use of the semiclassical approximation as shown above, we can predict that the maximum efficiency is around 0.9 and therefore the number of photons created will be $0.9\times(248+2\times251)=675$, simply by  analyzing the potential. 

\begin{figure}[t!]
	\begin{center}
		\includegraphics[width=8.9cm]{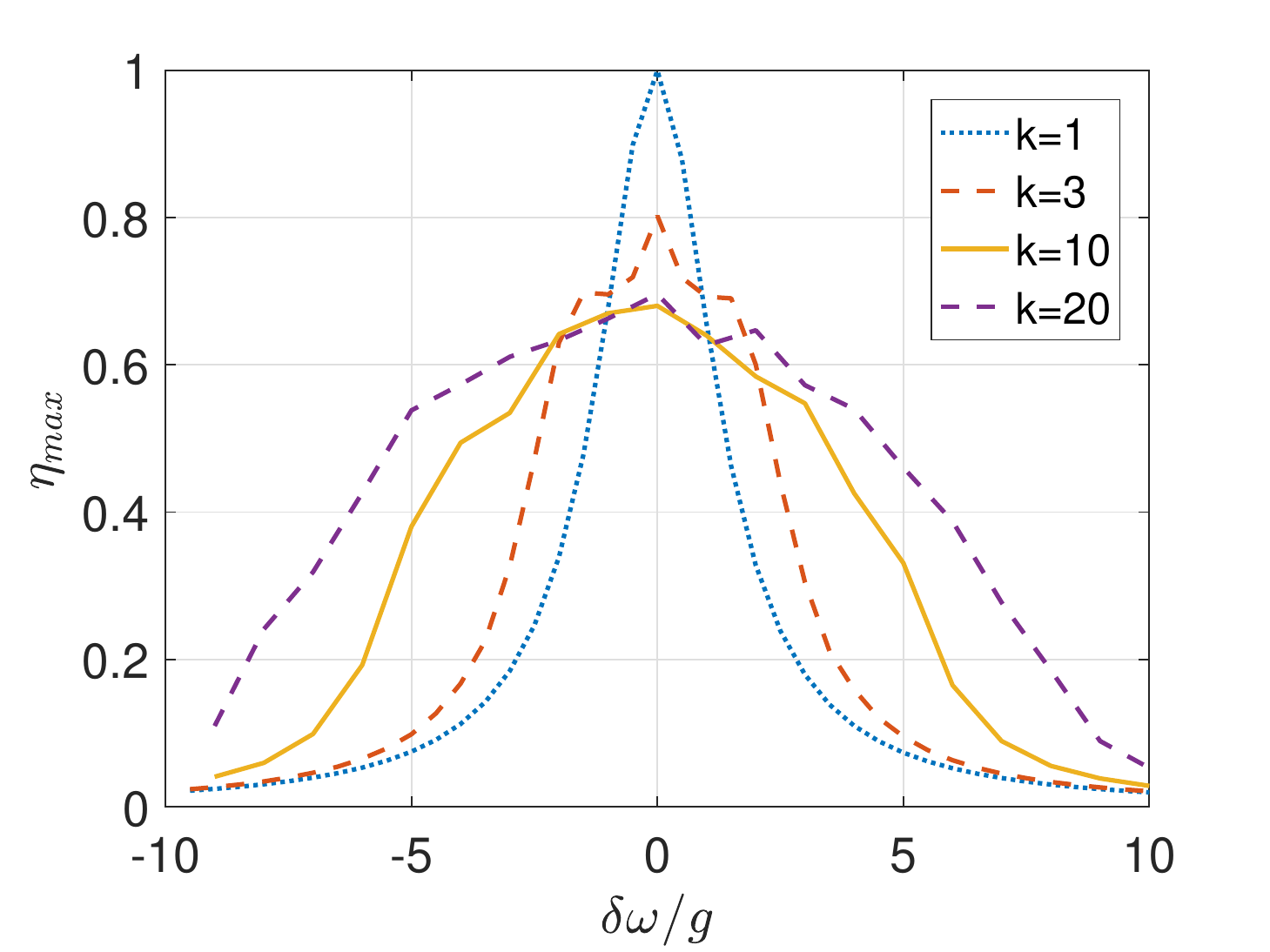}
		\caption{ Efficiency of energy conversion as a function of detuning for initial states of the form $|0,k\rangle$.}
		\label{fig:detune}
	\end{center}
\end{figure}

\subsection{Detuning}

We have shown that by adjusting the frequency of the mobile wall to twice that of the cavity it is possible to convert most of the energy of the system into DCE radiation. However, in real world experiments there is always  some detuning that reduces this effect. In this section, we study how precisely we need to tune these frequencies in order to observe photon emission.
\begin{figure*}
	\subfloat[\label{sfig:testa}]{%
		\includegraphics[width=.45\linewidth]{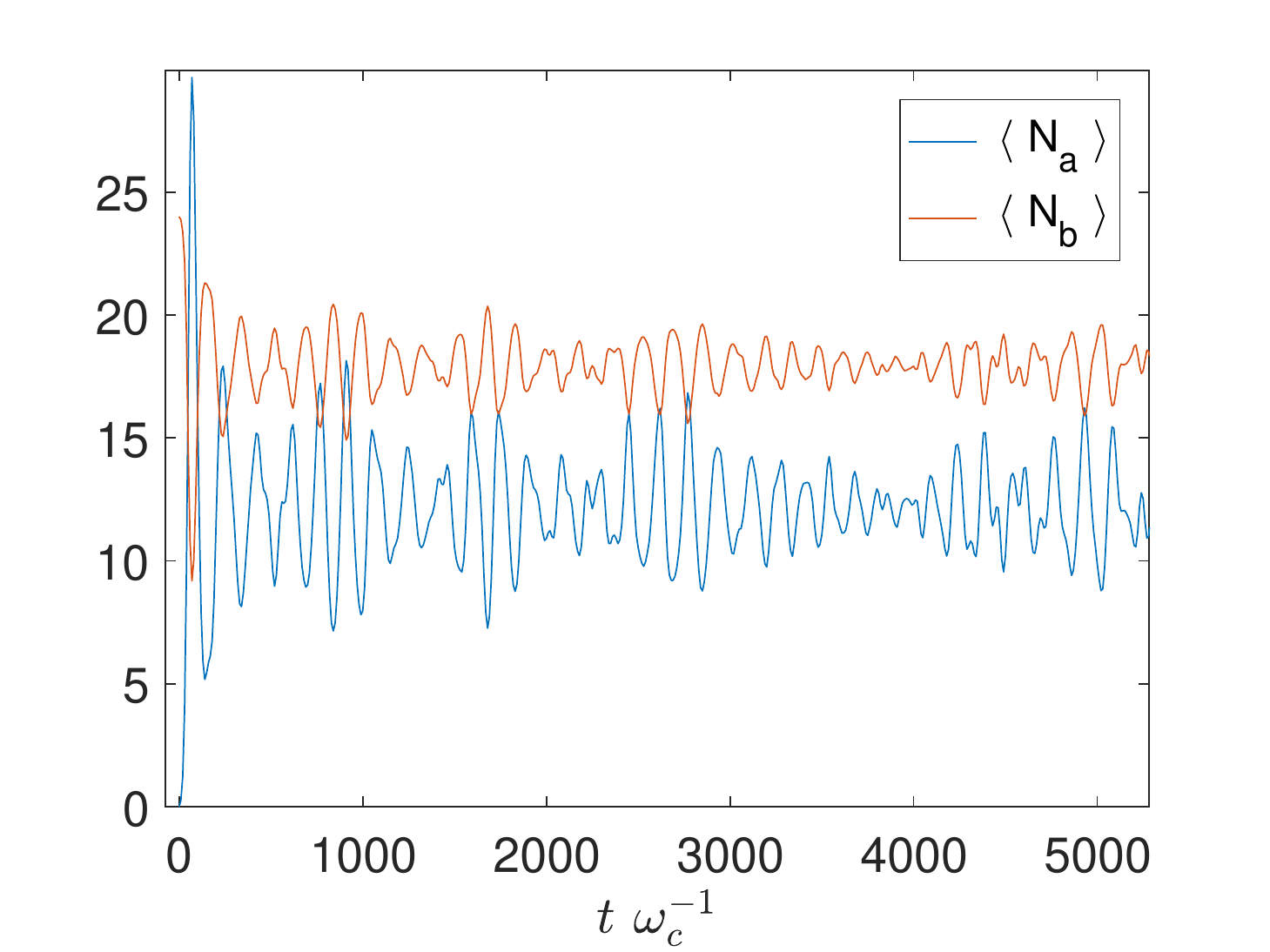}%
	}\hfill
	\subfloat[\label{sfig:testa}]{%
		\includegraphics[width=.45\linewidth]{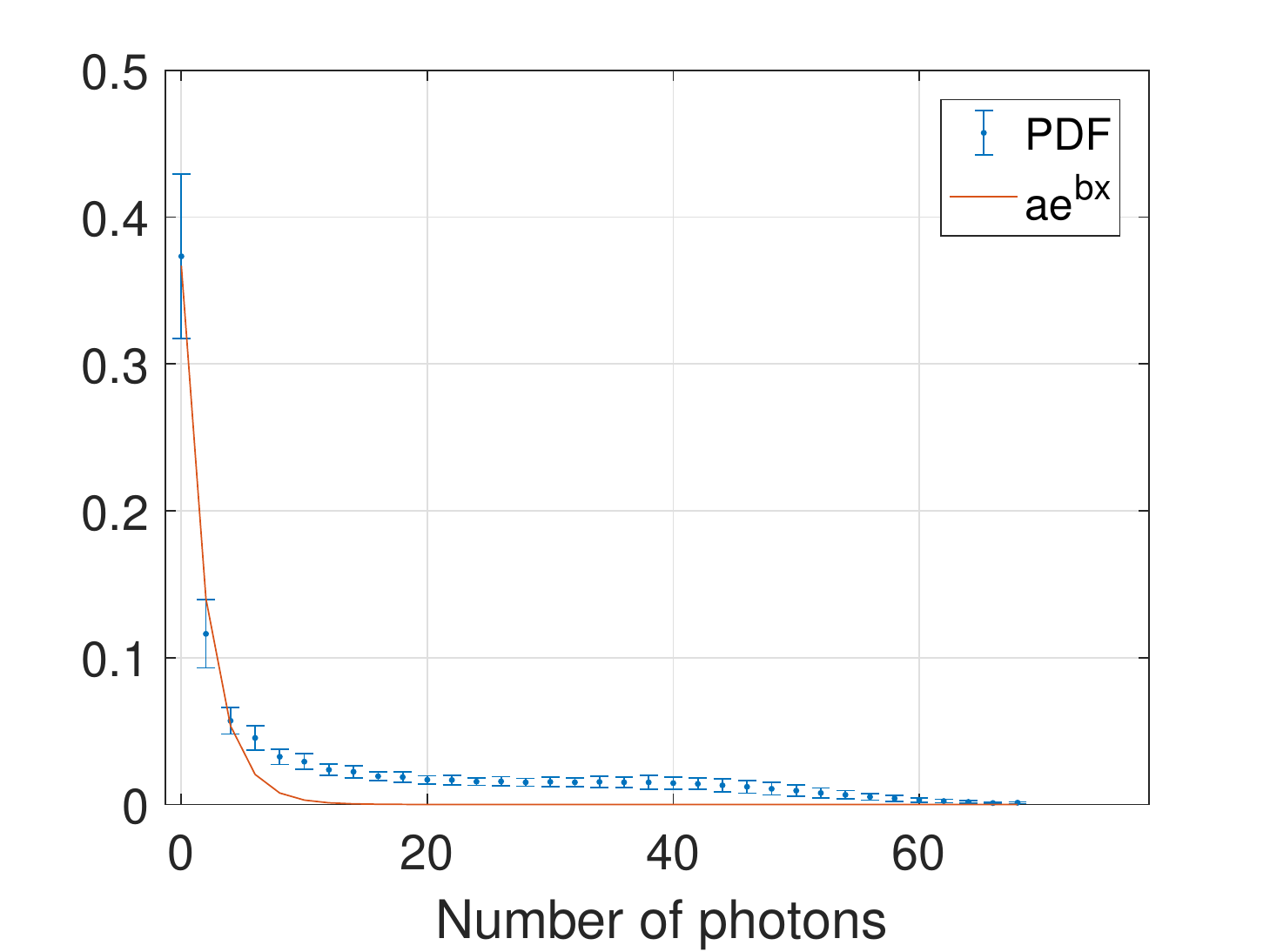}%
	}
	\caption{a) Time evolution of the mean photon and phonon number for the initial state $|0\rangle \otimes |\alpha=5\rangle$. b) Photons distribution with error bars indicating the magnitude of the time fluctuations and compared with an exponential fit.}
	\label{fig:coh_vm}
\end{figure*}
Firstly, we consider the initial state $|0,1\rangle$ and a mechanical frequency slightly detuned from parametric resonance $\omega_{\rm m}=2\omega_{\rm c}+\delta\omega$. We can solve the dynamics to the lowest order in perturbation theory by diagonalizing the new interaction 
\begin{equation}
	W_{\text{int}}=\frac{g\hbar}{2} V_{\text{DCE}}+\hat{N}_{ b}\delta\omega \hbar
\end{equation}
in the subspace $D_1$. In the basis $\left\{|0,1\rangle, |2,0\rangle \right\}$ the perturbation is
\begin{equation}
W_{\text{int}}\vert_{D_1}=\frac{g\hbar}{2}
\left(\begin{array}{cc}
2\delta\omega/g&\sqrt{2}\\
\sqrt{2}&0\\
\end{array}\right),
\end{equation}
and can be diagonalized to find the time evolved state 
\begin{equation}
	|\psi(t)\rangle=\cos(\Omega t)+\frac{i\delta\omega/g\sin(\Omega t)}{\sqrt{2+(\delta\omega/g)^2}}|0,1\rangle+\frac{i\sqrt{2}\sin(\Omega t)}{\sqrt{2+(\delta\omega/g)^2}}|2,0\rangle \nonumber
\end{equation}
with $\Omega=\sqrt{2}g\hbar/2$. This corresponds to a maximum efficiency of 
\begin{equation}
\eta_{\rm max}=\frac{\hbar\omega_{\rm c}\max_{t}\langle \hat{N}_{a}\rangle}{2\hbar\omega_{\rm c}}=\frac{2}{2+(\delta\omega/g)^2},
\end{equation}
which is a Lorentzian function that halves for $\delta\omega=g\sqrt{2}$. This means that even for $g=0.01$, we can still produce half of the resonance photons for a detuning of around $1.4\%$.

Further, we numerically studied how the efficiency depends with the detuning for various initial states of the form $|0,k\rangle$ finding that the width of the curve increases linearly with growing $k$ (Fig. \ref{fig:detune}). That is we have found that, even though the efficiency decreases in parametric resonance as we increase the energy, the photon production becomes less sensitive to the detuning of the cavity. This tell us that if we try to externally drive a cavity that is detuned we might not observe photon production until we have reached a critical amount of phonons in the wall.

\section{Coherent states}

As we have said, the DCE has been studied extensively in QFT by considering a classical wall oscillating harmonically in time. However, the initial states we have analyzed so far have no classical analogue since the position and momentum observables do not evolve in time. For that reason, in this section we study coherent states in order to describe the state of the phonons. Assuming that the cavity is initially in vacuum, we study the evolution of the state and the generation of photons.

The dynamics for an initially coherent phonon state of the form
\begin{equation}
	|\psi_0\rangle=|0\rangle \otimes |\alpha\rangle=e^{-|\alpha|^{2}/2}\sum_{k=1}^{\infty}\frac{\alpha^{k}}{\sqrt{k!}}|0,k\rangle
\end{equation}
differs greatly from those of the product basis since the time evolution of the mean number of photons shows an irreversible behavior. As shown in Fig. \ref{fig:coh_vm} (b), photons are produced rapidly, reaching to an all time maximum only to decrease to a stationary value for long times, $\bar{\langle N_{a} \rangle}$, around which they fluctuate. Even yet, the complete photon distribution becomes stationary showing a very high probability of measuring a small number of photons (less than the mean), followed by an exponential decay and then a plateau of equal probability for a large number of photons, shown  Fig. \ref{fig:coh_vm}(b). 
\begin{figure}[h!]
	\begin{center}
		\includegraphics[width=8cm]{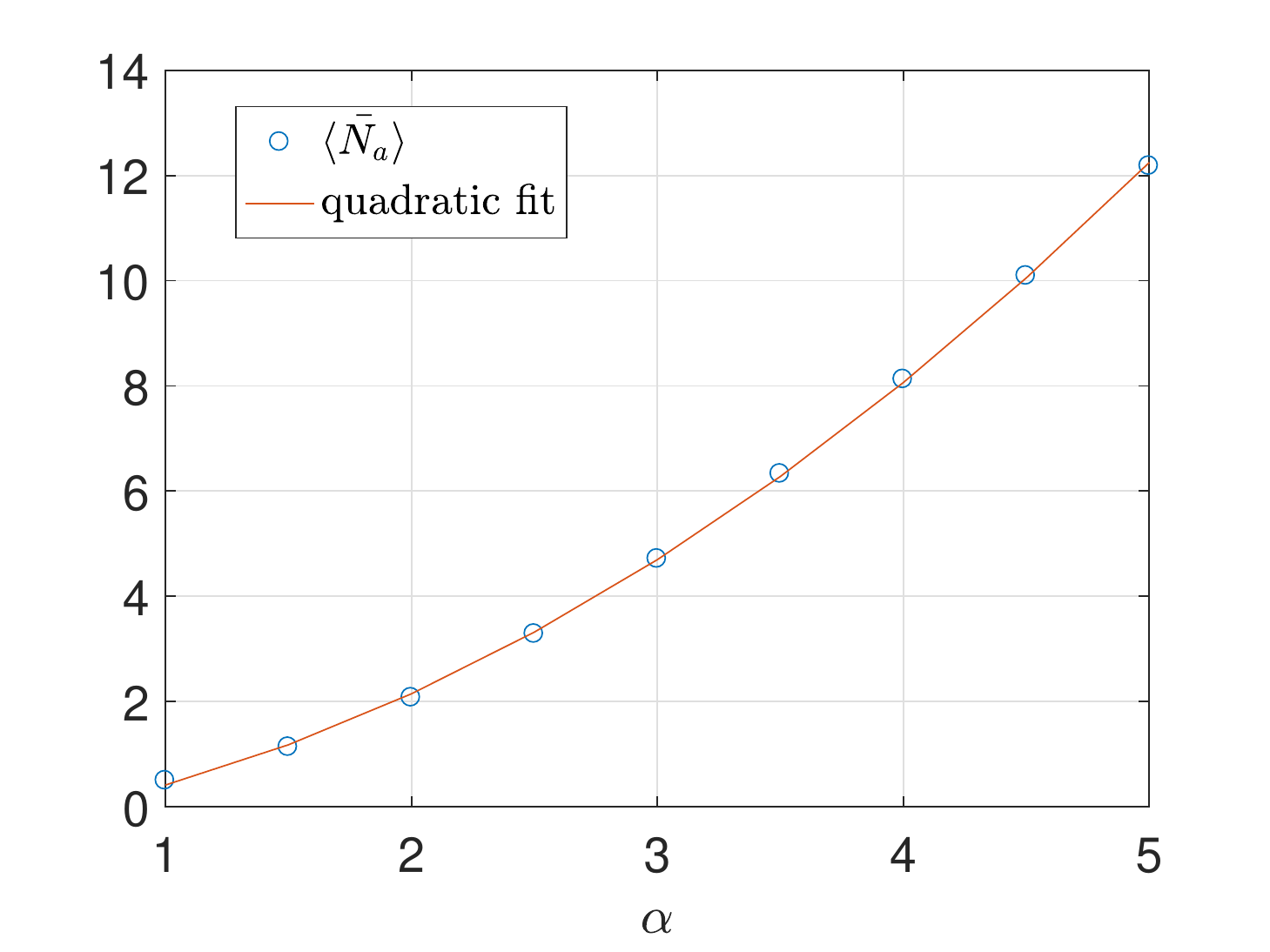}
		\caption{ The asymptotic mean photon number for initially coherent phonon states $|0\rangle \otimes |\alpha\rangle$ shows a quadratic increase with the initial displacement $\alpha$.}
		\label{fig:coh_alfa}
	\end{center}
\end{figure}
As for the stationary value of the mean photon number, it can be seen in Fig. \ref{fig:coh_alfa} that it actually grows quadratically with $\alpha$. It is possible to understand this behavior by  recalling the known result of DCE for a cavity with a moving wall of amplitude $\epsilon$. In such a case, it has been shown that the number of particles created out of vacuum by the movable mirror is
\begin{equation}
\langle \hat{N}_{a} \rangle=\sinh^2(\gamma \epsilon t),
\end{equation}
being $\bar{\langle N_a \rangle}\propto \epsilon^2$ for a small amplitude movement \cite{crocce}. This result is consistent with the result  from quantum field theory for the driven DCE with a classical wall as we obtain that the number of photons created  is quadratic in $\alpha$.
Finally, we have also seen that the efficiency  converges to around $25\%$ as we increase the amplitude $\alpha$. This can be understood with the results of the previous section. Since the initial state $|\psi_0\rangle$ is a superposition of many product states, each of which evolves independently in its own subspace with different frequencies, their oscillations compensate to give a constant number of mean photons. This number, for high value of $\alpha$, is given by the limit efficiency of product states for high energies corresponding to the mean efficiency of Fig. \ref{fig:tensor_eta_n} (for $n=0$ since we are starting with the cavity in the vacuum state).
On in all, the DCE interaction for coherent states can be seen as a sort of quantum friction that by converting phonons in the mirror to photons reduces the amplitude of the mirror oscillation. This has previously been noted in \cite{dalvit3} by a different technique tracing over the field degree of freedom and looking only at the mirror's motion. We can now understand the mechanism by which process occurs, quantify it and obtain the final state of the EM field. 
\begin{figure}[h!]
	\centering
	\includegraphics[scale=0.53]{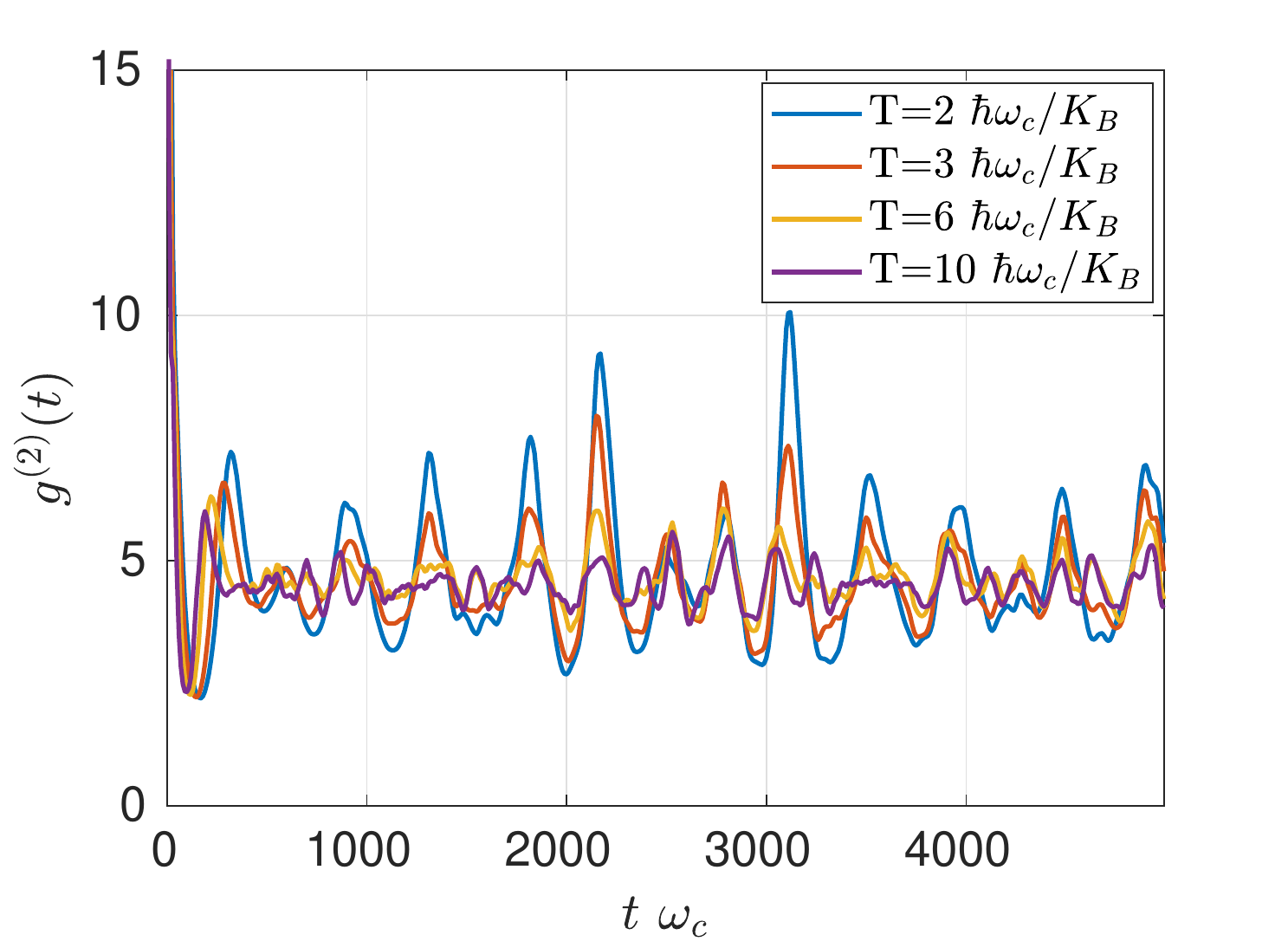}
	\caption{ The correlation function $g^{(2)}$ for photons takes values around $4.5$ which indicates super-poissonian statistics as would be expected for pair production of photons. The fluctuations around this values tend to diminish for higher temperatures. }
	\label{fig:g2}
\end{figure}


\section{Thermal states}
In this section, we shall consider another type of initial state for the phonons in the wall. As it can be technically very challenging the preparation of an initial coherent or tensor product state, we might consider a  much more accesible to prepare it in a thermal state. Thus, we herein study  the evolution of an initial state of thermal phonons in the wall and vacuum inside the cavity for a range of different initial temperatures, defined as  
\begin{equation}
	|\psi_0\rangle=\sum_{k=0}^{\infty}\frac{e^{-\hbar\omega_mk/(K_{\rm B}T_0)}}{\sqrt{1-e^{-2\hbar \omega_m/(K_{\rm B}T_0)}}}|0,k\rangle.
\end{equation}
We start by studying the equal-time photonic normalized second-order correlation
function for photons,
\begin{equation}
g^{(2)}(t,t)=\frac{\langle \hat{a}^\dagger(t) \hat{a}^\dagger(t) \hat{a}(t) \hat{a}(t) \rangle}{\langle \hat{a}^\dagger(t) \hat{a}(t) \rangle}, .
\end{equation}
shown in Fig.\ref{fig:g2}.
This magnitude reaches a stationary value for thermal states of around $4.5$ which tell us that the radiation produced has super-poissonian statistics and that the probability of producing two photons at once is higher than simple chance reenforcing the idea that photons are produced in pairs. As the temperature is increased the fluctuations around the stationary value become smaller since photon production becomes more uniform in time. 

\begin{figure*}
	\subfloat[\label{sfig:testa}]{%
		\includegraphics[width=.45\linewidth]{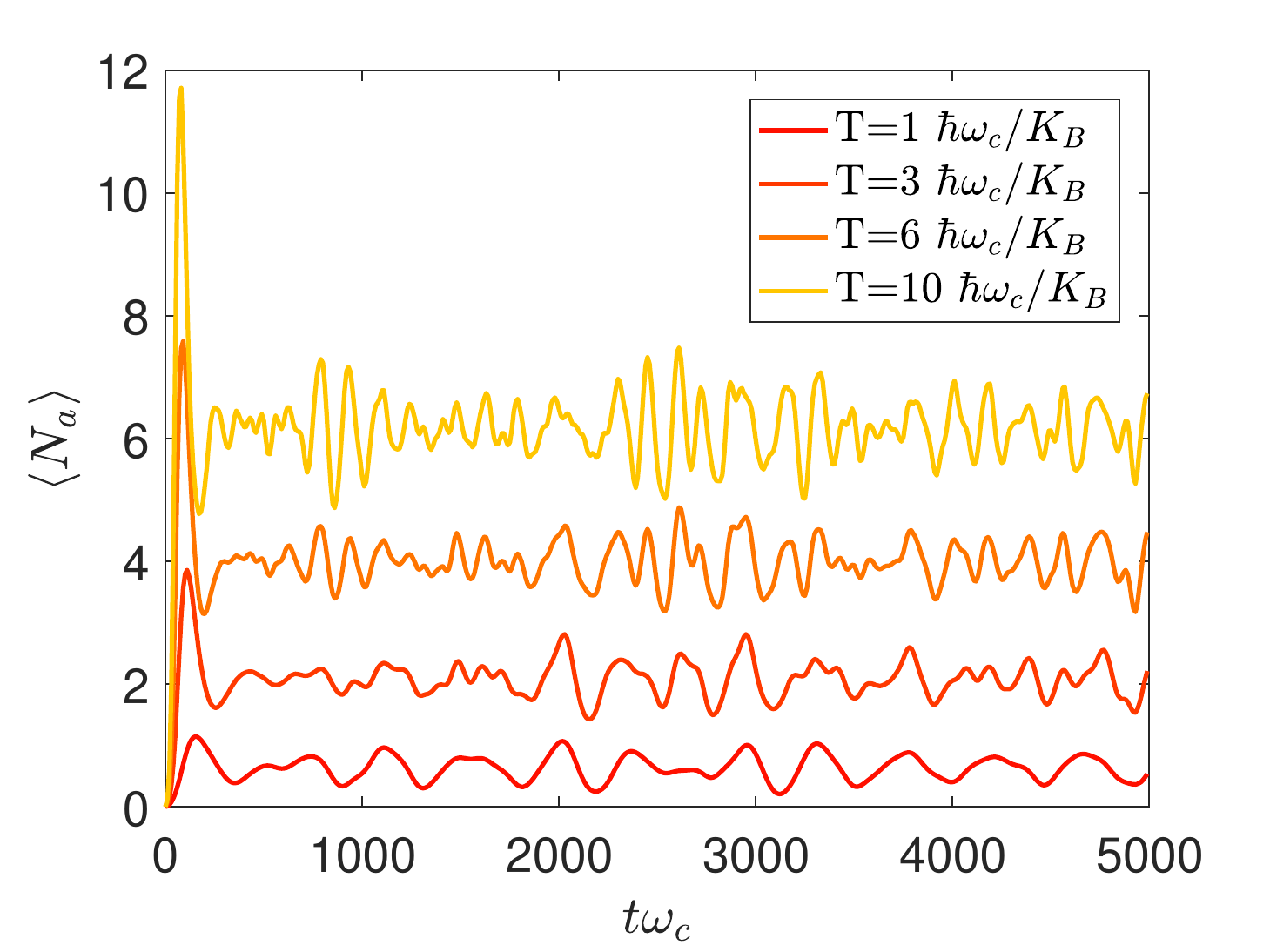}%
	}\hfill
	\subfloat[\label{sfig:testa}]{%
		\includegraphics[width=.45\linewidth]{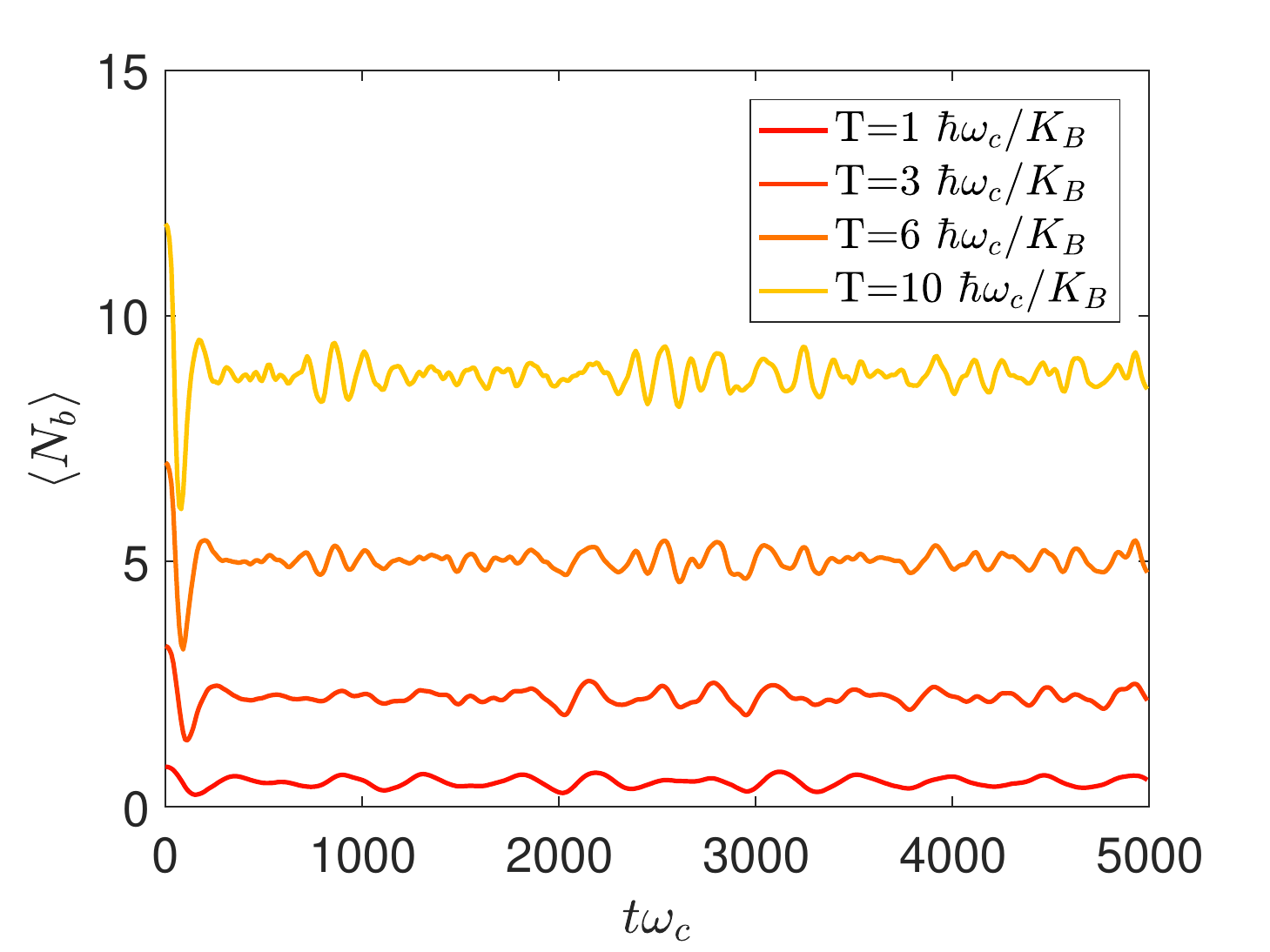}%
	}\hfill
	\subfloat[\label{sfig:termico_s_t}]{%
		\includegraphics[width=.45\linewidth]{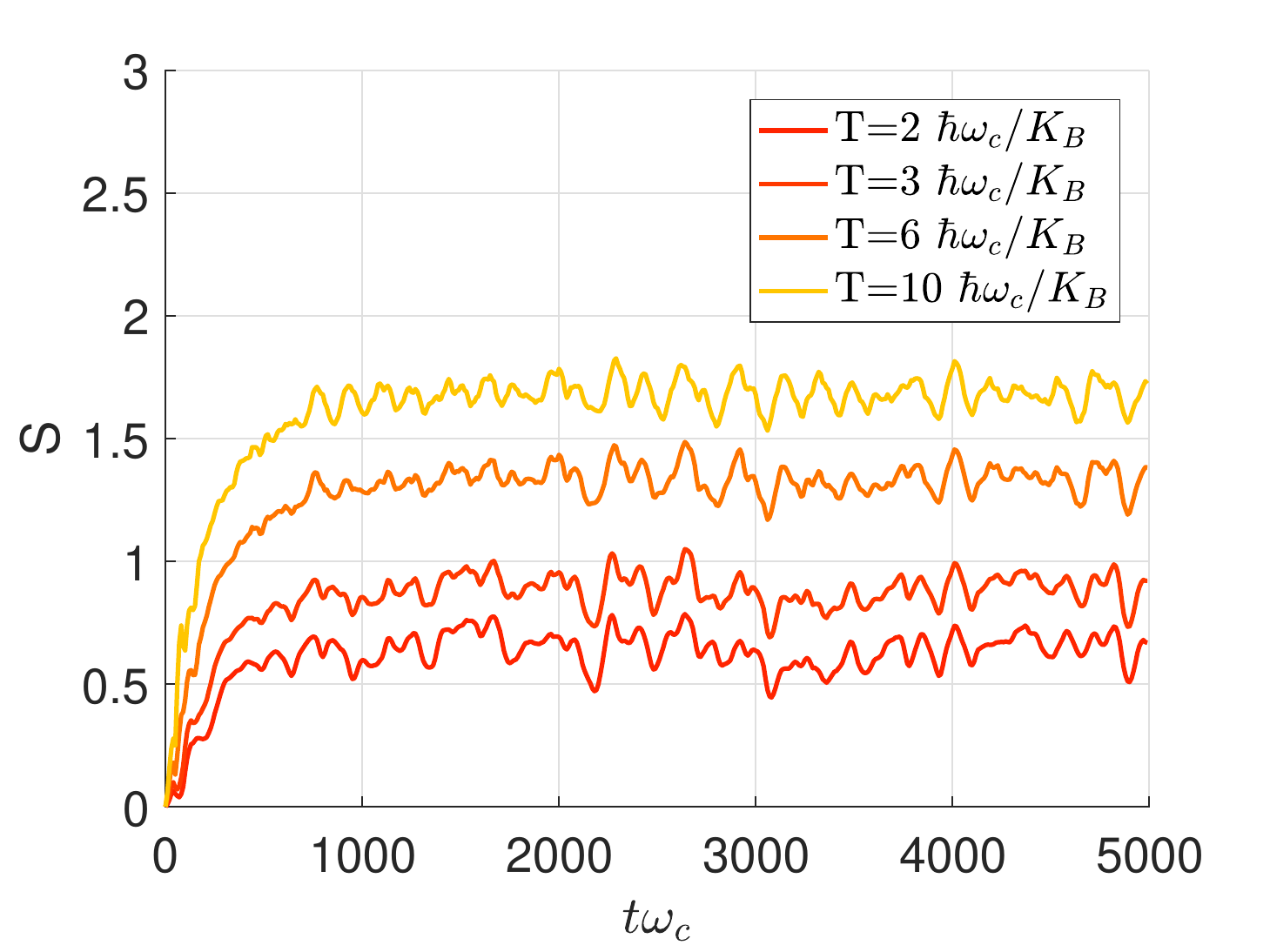}%
	}\hfill
	\subfloat[\label{sfig:termico_s_t_fit}]{%
		\includegraphics[width=.45\linewidth]{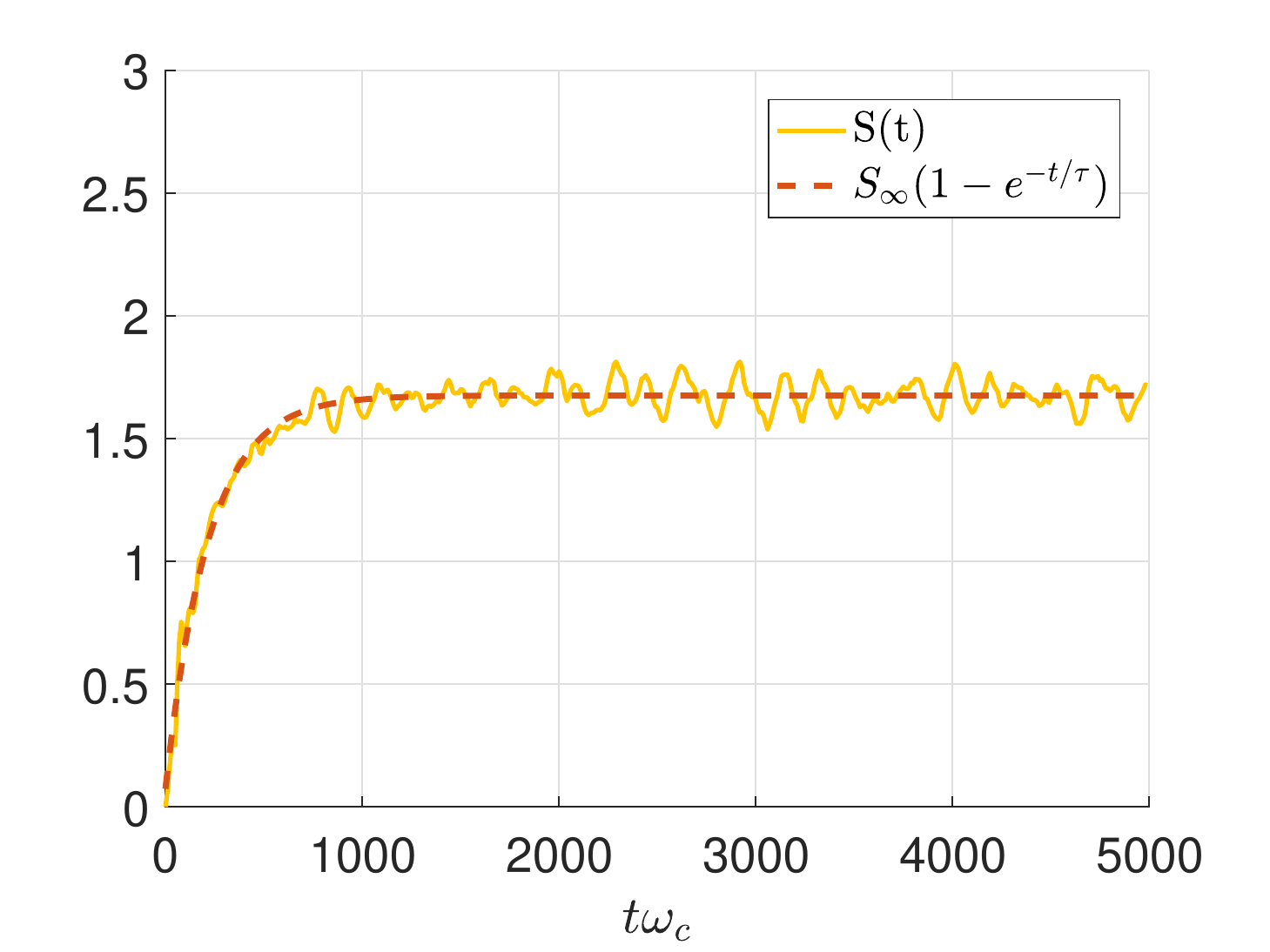}%
	}
	\captionsetup{justification=justified,singlelinecheck=off}
	\caption{a) Mean number of photons as a function of time for different temperatures. b) Mean number of phonons as a function of time for different temperatures. c) Entanglement entropy as a function of time of different temperatures. d) Entanglement entropy as a function of time for an initial state of phonons with an exponential fitting.}
	\label{fig:termico_vm}
\end{figure*}

Further, we continue by studying the number of photons created in the cavity as to compared with the different initial states considered above. In Fig. \ref{fig:termico_vm}(a), we present the mean value of photons as time evolves. We can easily detect a similar behavior to that of the coherent states: a  sudden increase of photons, followed by a very high peak and then a strong decrease to a stationary value around which there are fluctuations for long times. The dynamics of phonons is, of course, the opposite as to obey the conservation of energy due to the weak coupling regime Fig. \ref{fig:termico_vm}(b). Both observables increase with higher temperatures and their fluctuations decrease with respect to their stationary mean. 
On the other hand, the entanglement entropy  increases with time and reaches a stationary value for long times as shown in Fig. \ref{fig:termico_vm}(c). In this case, it becomes evident that the equilibrium occurs for times bigger than $10^3\ \omega_{\rm c}^{-1}$ for all the temperatures considered. Even more so, it is possible to fit the entanglement evolution with the function
\begin{equation}
	S(t)=S_\infty(1-e^{-t/\tau}),
\end{equation}
where $S_\infty$ and $\tau$ are constants, to a very high degree of accuracy, as shown in (Fig. \ref{sfig:termico_s_t_fit}).

\subsection{Distributions and thermalization}
\begin{figure*}
	\subfloat[\label{sfig:termico_dist_n}]{%
		\includegraphics[width=.45\linewidth]{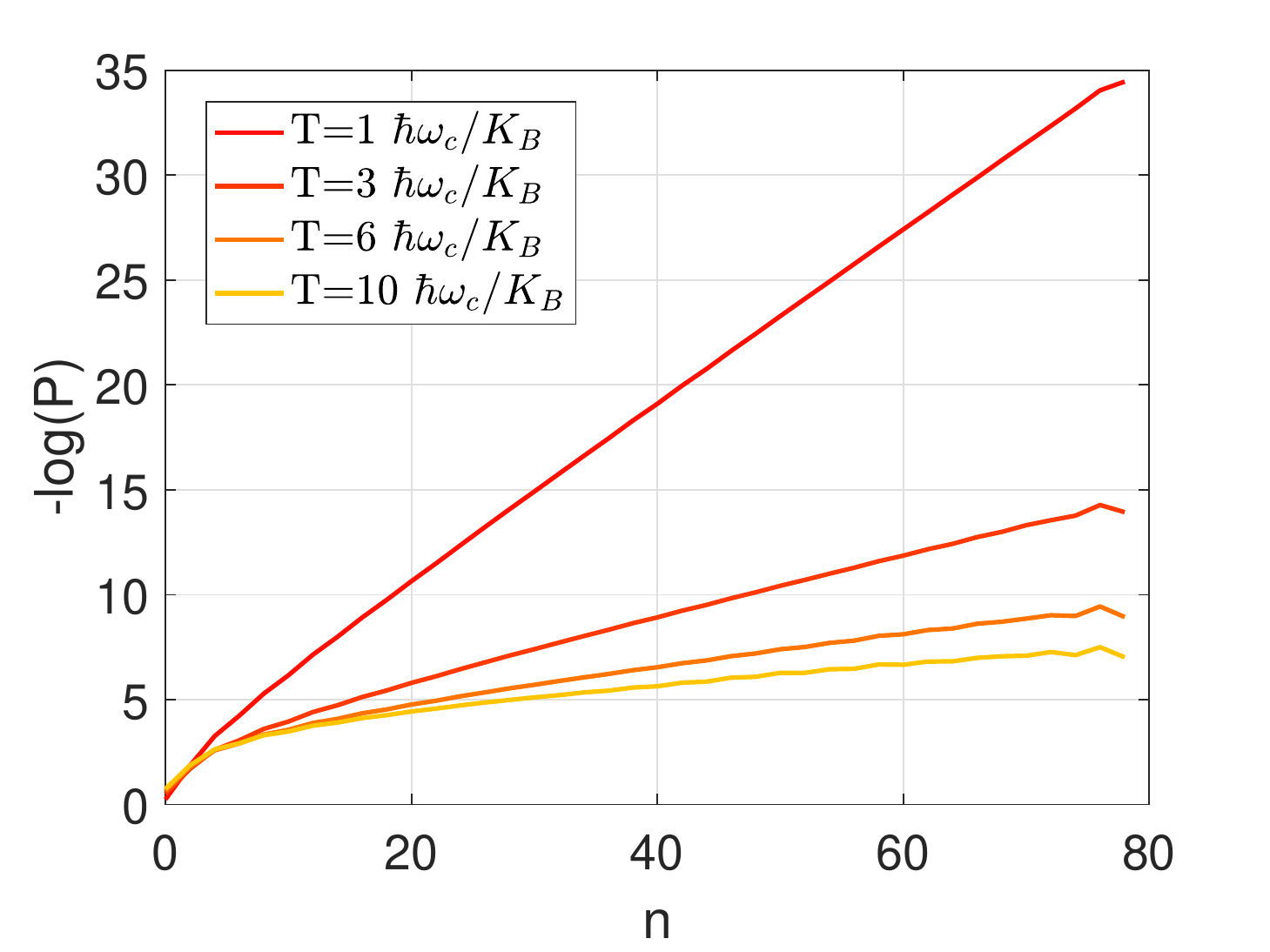}%
	}\hfill
	\subfloat[\label{sfig:testa}]{%
		\includegraphics[width=.45\linewidth]{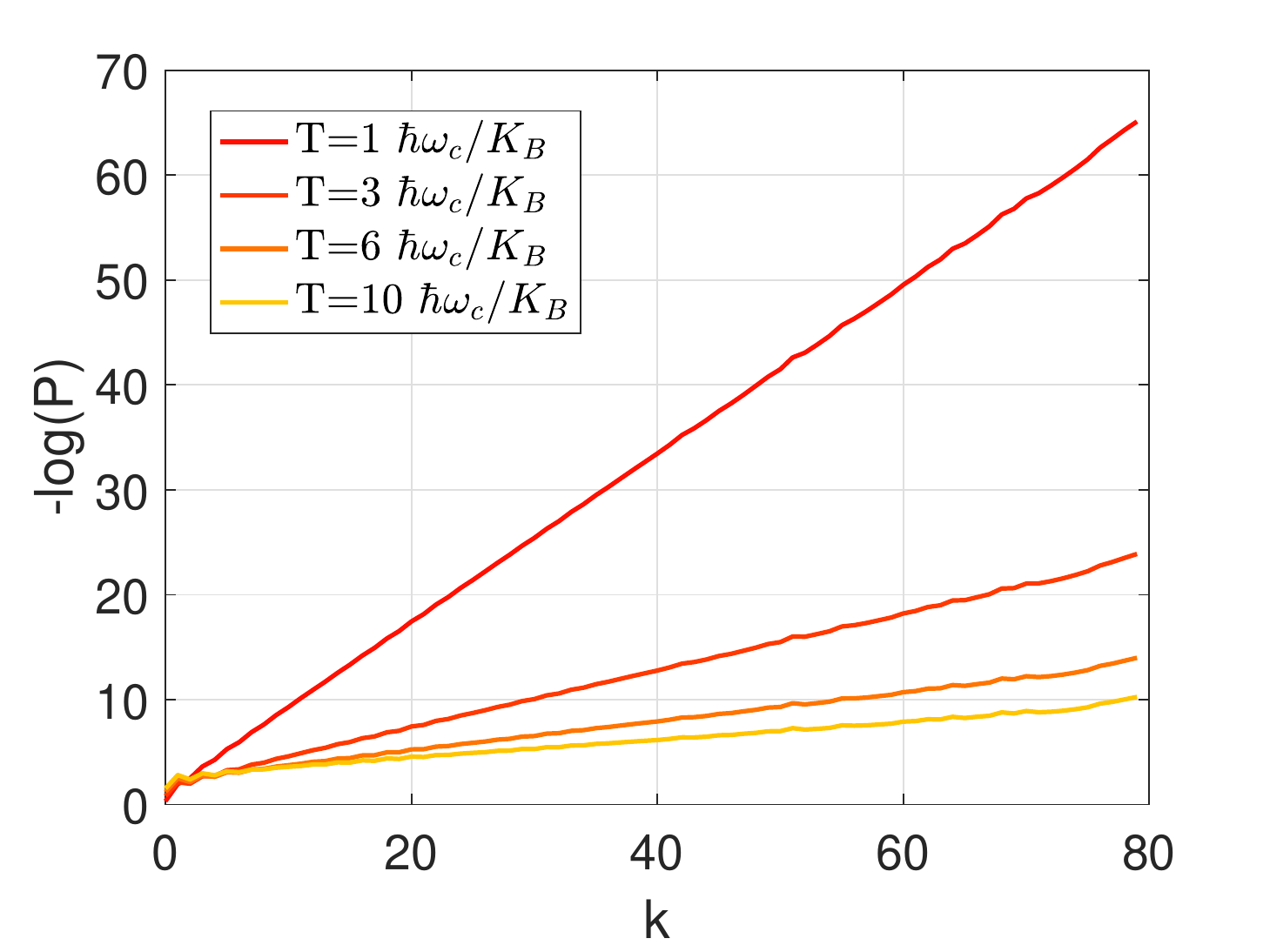}%
	}\hfill
\subfloat[\label{sfig:termico_beta2}]{%
	\includegraphics[width=.45\linewidth]{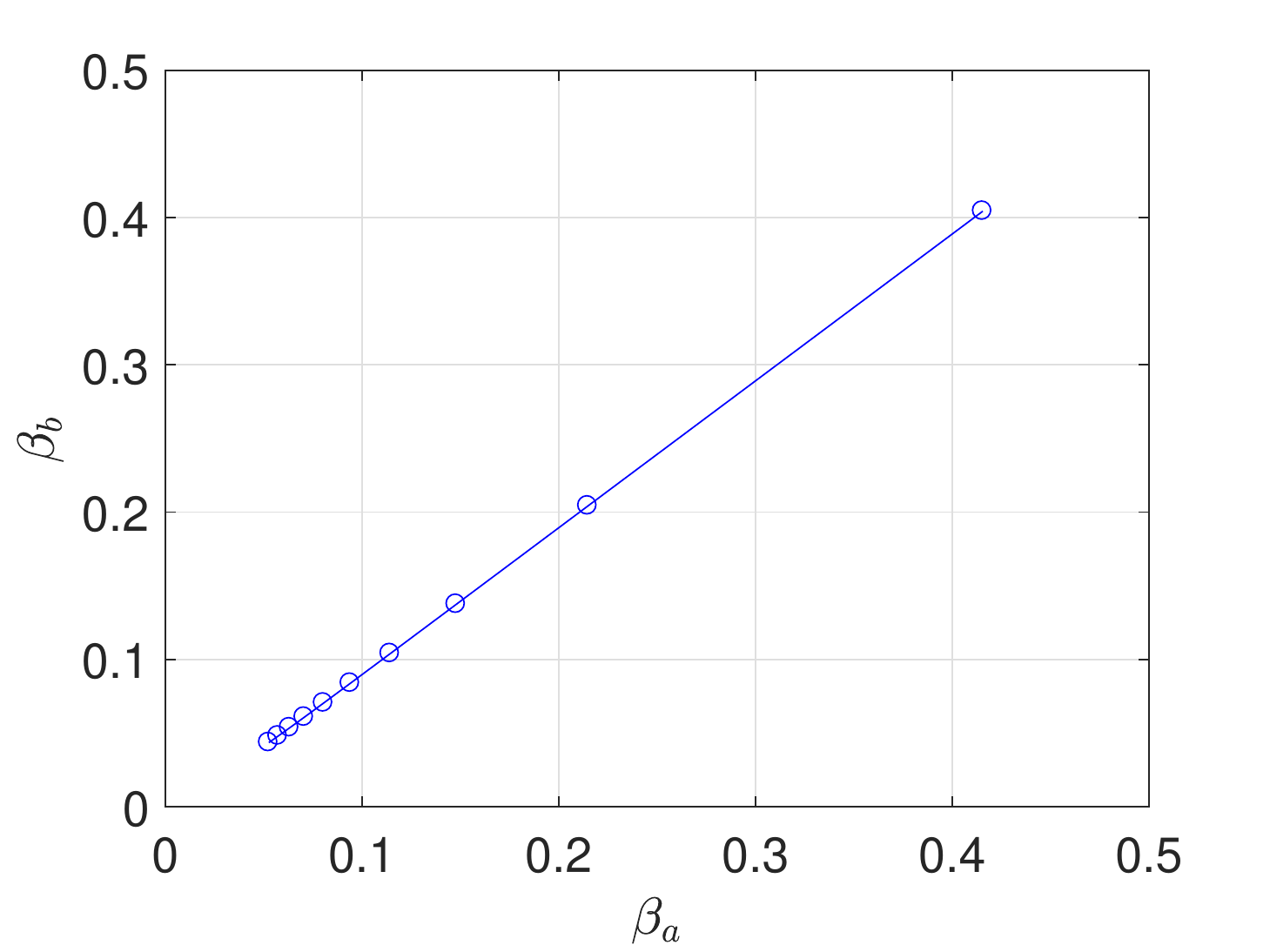}%
}\hfill
\subfloat[\label{sfig:termico_eta}]{%
	\includegraphics[width=.45\linewidth]{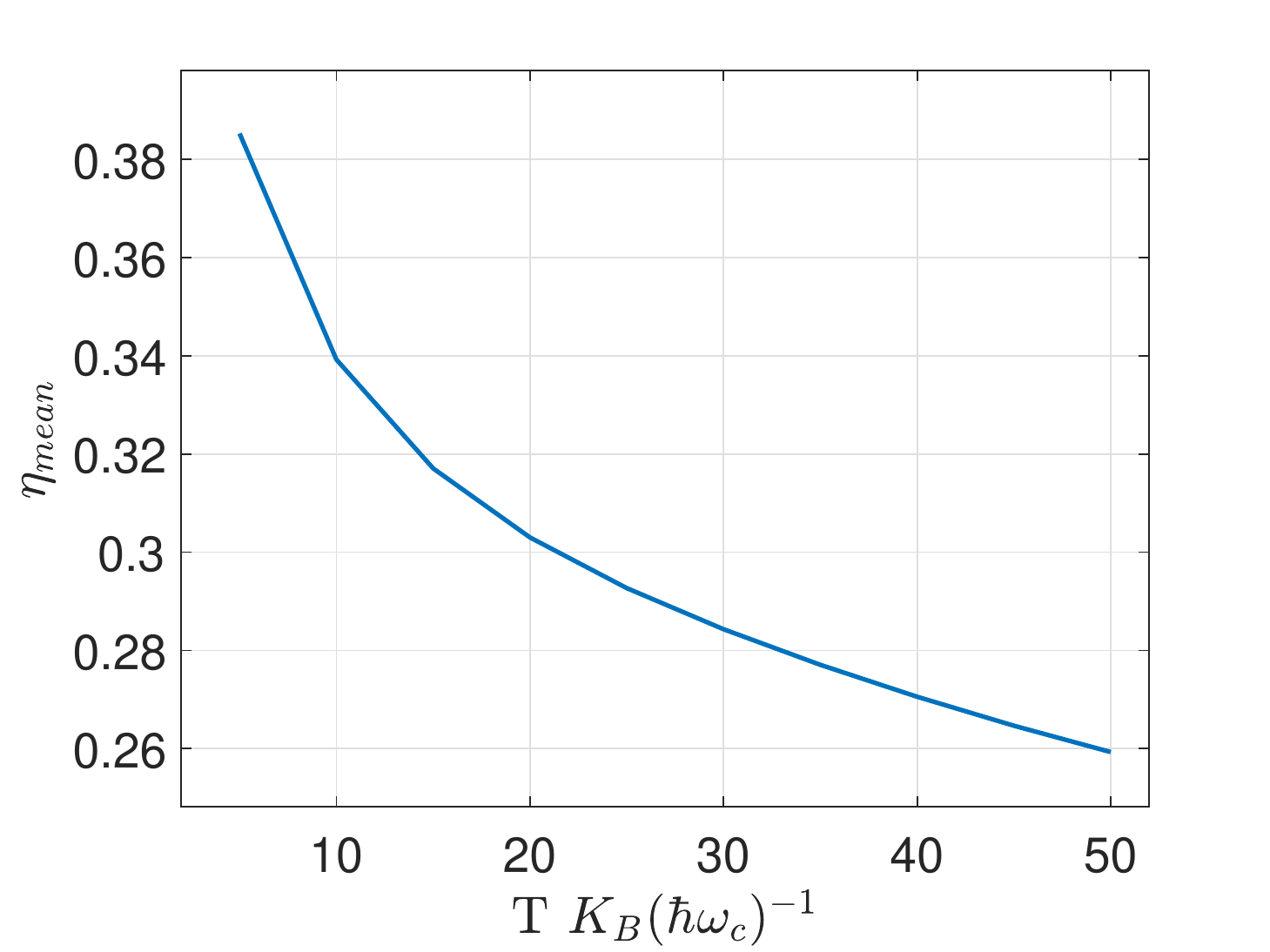}%
}
	\caption{ Logarithmic distribution of photons (a) and phonons (b) for different initial temperatures. The slope which is proportional to inverse temperature reduces as the initial temperature increases. c) The relationship between the photon and phonon temperatures is linear with slope 1, which indicates that the subsystems share the same temperature. d) Efficiency of thermal states with different temperatures.}
	\label{fig:termico_dist}
\end{figure*}
The fact that the mean number of photons and phonons reach a stationary value and that the entropy seems to saturate for long times suggests that the system reaches a stationary state. In this subsection, we try to characterize the stationary state attained by looking at the photon and phonon distributions for different initial temperatures and the relationship among them. We consider the photon (respectively phonon) distribution $P_{\rm \rho}(n)$ of a state $\hat{\rho}(t)$ to be the diagonal elements of the photon (respectively phonon) reduced density matrix in the energy basis. It can be seen that the off-diagonal elements are negligible and are not necessary to reproduce the expectation values after equilibration. 
Given the initial thermal state in the wall, we might suspect that a thermalization process is occurring. If that was the case, then the reduced photon matrix should be given by a Gibbs state
\begin{equation}
	\hat{\rho}_{a}=\sum_{n\ \text{even}} \frac{e^{-\beta_{a} H_{a}}}{Z}|n\rangle \langle n|=\sum_{n\ \text{even}} \frac{e^{-\beta_{a} \hbar \omega_{\rm c} n}}{Z}|n\rangle \langle n|,
\end{equation}
where $\beta_{a}=1/(K_B T_a)$ is the inverse photon temperature and $Z$ the partition function.
Likewise, the  photon distribution would be of the form
\begin{equation}
	P(n)=\frac{e^{-\beta_{a} \hbar \omega_{\rm c} n}}{Z}.
\end{equation}
In Fig. \ref{fig:termico_dist}, we present $-\log(P(n))$ for different initial values of $n$ and different values of the initial temperature for photons  (Fig. \ref{sfig:termico_dist_n}). We can see that the behavior fits an almost perfectly straight line for large $n$ and for all temperatures, with a decreasing slope as the initial temperature increases. The same behavior can be reproduce for phonons (Fig. \ref{sfig:testa}), suggesting that both subsystems can be extremely close to a thermal state.

As we have further evidence that a thermalization process might have taken place at this stage, we might consider  defining a temperature for the whole system. In the case of a thermal state the (inverse) temperature would be given by the slope of the function 
\begin{equation}
	-\log(P(n))= \beta_{\rm a} \hbar \omega_{\rm c} n+\log(Z).
\end{equation}
By fitting a straight line for this magnitude for photons and phonons we are able to define an inverse temperature $\beta_{a}$ and $\beta_{b}$ for both subsystems. By studying the relationship between these two temperatures for a range of initial states, we find that they are indeed the same and so we can say that the system thermalizes, as shown in Fig. \ref{sfig:termico_beta2}. 

Thermal states can be much easier to prepare while not being too inefficient. In fact their mean efficiency is around $25\%$ for high temperatures (Fig. \ref{sfig:termico_eta}), which is approximately the same as that for initial product states with a high number of photons (compared to Fig. \ref{fig:tensor_eta_n} for $n=0$).

\section{Conclusions}\label{sec:conc}

In this paper we have studied the dynamical Casimir effect as a closed quantum system described as an interaction between photons in a cavity and phonons in a moving wall. We have found that the efficiency of photon production reduces as the energy increases for initial states of the form $|0,k\rangle$, approaching asymptotically around $70\%$. The entanglement entropy, both in mean and maximum value, on the other hand increases logarithmically with the energy and are related by $S_{\rm mean}=0.7\ S_{\rm max}$. We have also seen that by starting the evolution with a few photons already in the cavity we can actually stimulate the emission of more photons, increasing the efficiency. However, we have shown that, if we keep adding more initial photons, the efficiency linearly decreases inhibiting photon generation. In fact, we have a stable equilibrium where if the initial state is of the form $|n,k\rangle$ with $0.56k\leq n \leq 0.66k$ there is almost no photon or phonon production. We were also able to obtain a differential equation for the number of photons indicating that it evolves in time just as a non-harmonic quantum oscillator.
The dependence of detuning for the DCE was also studied, finding a Lorentzian curve of photon production with a width proportional to the coupling of the system and the energy of the initial state. That is the DCE is less sensitive to the detuning as we increase the energy of the initial state.

The dynamics found for coherent states $|0, \alpha\rangle$ is very different from that of the product basis, since the mean number of photons and phonons seem to reach a stationary value. We have found that these values depend quadratically with the parameter $\alpha$ of the initial state, which seems to indicate a quadratic dependence of the photon production with the amplitude of the oscillation of the wall. This result is  consistent with  QFT results derived in previous studies found in the literature.

Finally we have analyzed the time evolution of an initial thermal state in phonons and vacuum in the cavity. In this case both the number of phonons and photons seem to reach a stationary value. Similarly, the entanglement is found to be very well described by a function of the form $S(t)=S_{\infty}(1-e^{-t/\tau})$, saturating for times longer than $10^3 \omega_{\rm c}^{-1}$ for all initial temperatures. We have also seen that the stationary state reached by the system is very close to a thermal one for both photons and phonons. Even more so, we have been able to define a temperature for both subsystems and show that it is actually the same for different initial states which leads us to conclude that the system finally thermalizes. 

Several new lines of research arise from this work. It would be interesting to look at how these results change outside of parametric resonance, with the phonon frequency in some other Casimir-Rabi splitting. Another possible direction would be to study the system's response in the strong coupling regime where the counterrotating terms become relevant. Coming back to the root of the interaction at hand, it would be important to analyze how the driving of the mirror modifies the behavior and compare it with previous dynamical Casimir  effect results in the context of quantum field theory with a classical wall. Finally, our model describes a typical three dimensional cavity with a non equidistant spectrum where only one mode of the EM field can be excited; however, in an one dimensional cavity, the spectrum is equidistant and many more modes can be excited. Hence, it would be relevant to study a new model where more modes can couple to each other.
\section*{Acknowledgements}
This work was supported by ANPCyT, CONICET, and Universidad de Buenos Aires - Argentina. FCL acknowledges International Centre for Theoretical Physics and 
Simons Associate Programme.

\end{document}